\documentclass[twocolumn,showpacs,showkeys,preprintnumbers,amsmath,amssymb,pra,aps,10pt]{revtex4-2}
\usepackage[T1]{fontenc}
\usepackage{color}
\usepackage{soul}
\usepackage{graphicx}
\usepackage{dcolumn}
\usepackage{bm}
\usepackage[colorlinks=true,linkcolor=blue,citecolor=red]{hyperref}
\newcommand{\sizetwo}{0.485\textwidth}

\usepackage{color}

\begin{document}

\preprint{\emph{Submitted to:} Physica A: Statistical Mechanics and its Applications}

\title{Thermodynamic and electromagnetic properties of \\ the eta--pairing superconductivity in the 
Penson--Kolb model}%

\author{Wojciech R. Czart}
\email[\mbox{e-mail: }]{czart@amu.edu.pl}
\homepage[\mbox{ORCID ID: }]{https://orcid.org/0000-0002-5859-1646}
\affiliation{Faculty of Physics, Adam Mickiewicz University in Pozna\'{n},
ulica Uniwersytetu Pozna\'{n}skiego 2, PL-61614 Pozna\'{n}, Poland}

\author{Konrad J. Kapcia}
\email[\mbox{corresponding author; e-mail: }]{konrad.kapcia@amu.edu.pl}
\homepage[\mbox{ORCID ID: }]{https://orcid.org/0000-0001-8842-1886}
\affiliation{Faculty of Physics, Adam Mickiewicz University in Pozna\'{n},
ulica Uniwersytetu Pozna\'{n}skiego 2, PL-61614 Pozna\'{n}, Poland}

\author{Roman Micnas}
\email[\mbox{e-mail: }]{rom@amu.edu.pl}
\homepage[\mbox{ORCID ID: }]{https://orcid.org/0000-0003-1574-8049}
\affiliation{Faculty of Physics, Adam Mickiewicz University in Pozna\'{n},
ulica Uniwersytetu Pozna\'{n}skiego 2, PL-61614 Pozna\'{n}, Poland}

\author{Stanis\l{}aw Robaszkiewicz}
\email[Deceased on 7th June 2017]{\ }
\affiliation{Faculty of Physics, Adam Mickiewicz University in Pozna\'{n},
ulica Uniwersytetu Pozna\'{n}skiego 2, PL-61614 Pozna\'{n}, Poland}

\date{\today}

\begin{abstract}
In the paper, we study the thermodynamic and electromagnetic properties of the Penson--Kolb (PK) model, i.e., the tight--binding model for fermionic particles with the pair-hopping interaction $J$.
We focus on the case of repulsive $J$ (i.e., $J<0$), which can stabilize the
eta-pairing superconductivity with Cooper-pair center-of-mass momentum 
$\vec{q}=\vec{Q}$, $\vec{Q}=(\pi/a$,$\pi/a$,\ldots). 
Numerical calculations are performed for several $d$-dimensional hypercubic lattices: 
$d=2$ (the square lattice, SQ), $d=3$ (the simple cubic lattice) and $d=\infty$ hypercubic lattice (for arbitrary particle concentration $0<n<2$ and temperature $T$).
The ground state $J$ versus $n$ phase diagrams and the crossover to the Bose--Einstein condensation regime
are analyzed and  the evolution of the superfluid characteristics are examined within the (broken symmetry) 
Hartree--Fock approximation (HFA). 
The critical fields, the coherence length, the London penetration depth, and the Ginzburg ratio are determined at $T=0$ and $T>0$ as a function of $n$ and pairing strength. 
The analysis of the effects of the Fock term on the ground state phase boundaries and on selected PK model characteristics is performed as well as the influence of the phase fluctuations on the eta-pairing superconductivity is investigated.
Within the Kosterlitz--Thouless scenario, the critical temperatures $T_{KT} $ are estimated for $d=2$ SQ lattice and compared with the critical temperature $T_c$ obtained from HFA. 
We also determine the temperature $T_m$ at which minimal gap between two quasiparticle bands vanishes in the eta-phase.  
Our results for repulsive $J$ are contrasted with those found earlier for the PK model with attractive $J$ (i.e., with $J>0$). 
\end{abstract}

\keywords{Penson-Kolb model, unconventional superconductivity, eta-pairing, phase diagrams, Kosterlitz--Thouless scenario, nonlocal pairing mechanism\\
\textbf{Highlights:}\\
\begin{itemize}
\item	The eta-pairing superconductivity in the Penson-Kolb model is studied. 
\item	BCS-BEC crossover for hypercubic lattices is investigated at $T=0$.
\item	Critical fields, coherence length, penetration depth are found for the ground state.
\item	The ranges of eta-pairing phase occurrence are determined for $T=0$ and $T>0$.
\item	The Kosterlitz-Thouless temperature for square lattice is estimated.
\end{itemize}
}

\maketitle

\section{Introduction}
\label{sec:intro}

The aim of the present work is to study superconducting properties of the Penson--Kolb (PK) model, i.e., the tight--binding lattice model with intersite pair hopping term $J$ (charge-exchange interaction)~\cite{Penson-86,Robaszkiewicz-99,Sikkema-95a,Sikkema-95,Bossche-96,Bouzerar-97,Japaridze-97,Japaridze-00,Roy-97,czart-11,czart-14,czart-19,Dolcini-00}.
This nonlocal pairing mechanism, which is a driving force of pair formation and their condensation, is distinct from the local one described by the attractive Hubbard (AH) model.
The pair-hopping term $J = \langle ii| e^2/r |jj \rangle$ can be derived  from the general tight-binding 
Hamiltonian \cite{Hubbard-63,Hubbard-63a,Kapcia-15,Kapcia-15a,KapciaJPSJ2016} selecting relevant terms of the two-particle interaction. 
Originating from the two-body potential, the site off-diagonal pair-hopping term describes part of the so-called ``bond-charge'' interaction.

The PK model is one of the simplest effective models for studying phenomenon of superconductivity, particularly in systems with almost unretarded and very short--ranged pairing. 
The Hamiltonian of the model investigated in this work has the following form: 
\begin{eqnarray}
\hat{H}& = &
-t\sum_{\langle i,j \rangle,\sigma }\left( {e^{\mathbf{i}\Phi _{ij}}\hat{c}_{i\sigma }^{\dag} \hat{c}_{j\sigma
}^{\ }+h.c.}\right)  \label{H} \\  
& - &{\frac{1}{2} J \sum_{\langle i,j \rangle}
\left( e^{2\mathbf{i}\Phi _{ij}} {\hat{c}_{i\uparrow}^{\dag} \hat{c}_{i\downarrow }^{\dag} \hat{c}_{j\downarrow
}^{\ } \hat{c}_{j\uparrow }^{\ } }+h.c.\right) }  
- \mu \sum_{i,\sigma } \hat{c}_{i\sigma }^{\dag} \hat{c}_{i\sigma}^{\ },
\nonumber
\end{eqnarray}
where parameters $t$, $J$, and $\mu$ denote the single particle hopping integral, the pair hopping (intersite charge exchange interaction), and the chemical potential, respectively. 
$\langle i, j \rangle$ restricts the summation to nearest neighbors (NN), independently.
The Peierls factors in Eq. (\ref{H}) take into account the coupling of electrons to the magnetic field via its vector potential $\vec{A}(\vec{r})$: $\Phi _{ij}=-\frac{e}{\hbar c}\int_{\vec{R}_i}^{\vec{R}_j} d\vec{r}\vec{A}(\vec{r})$ (where $e$ denotes the electron charge).

For attractive $J$ (i.e., $J > 0$) the system develops $s$-wave pairing states with total momentum $\vec{q}=0$ and corresponding order parameter $x_0=(1/N)\sum_i \langle \hat{c}_{i\uparrow}^\dag \hat{c}_{i\downarrow}^\dag \rangle \neq 0$
\cite{czart-10,czart-11,czart-19}. 
In the present work, we focus on the case of repulsive $J$ ($J<0$), which favours eta--pairing superconductivity with Cooper-pair center-of-mass momentum $\vec{q}=\vec{Q}$, $\vec{Q}= (\pi/a,\pi/a, \ldots)$ ($\vec{Q}$ is a half of the largest reciprocal lattice vector in the first Brillouin zone), and with order parameter defined as $x_{\eta}=(1/N)\sum_i \exp\left(\mathbf{i} \vec{Q} \cdot \vec{R}_i \right) \langle \hat{c}_{i\uparrow}^{\dag} \hat{c}_{i\downarrow}^{\dag} \rangle\neq 0$~\cite{czart-14,czart-15,czart-19}. 
Note that both these superconducting states are formally different types of superconductivity with the simplest isotropic pairing.

It is assumed that intercation parameters $t$ and $J$ are effective and they include all possible renormalizations and contributions such as, e.g., those coming from the coupling between electrons and other electronic subsystems, or those associated to the strong electron-phonon couplings in solid or chemical complexes \cite{Micnas-90} such as intermolecular  vibrations via modulation of the hopping integral \cite{Fradkin-83,Fradkin-83a} or from the on-site hybridization term in a generalized periodic Anderson model \cite{Robaszkiewicz-87n,Bastide-88}.

Model (\ref{H}) has been investigated only in several particular limits 
\cite{Penson-86,Robaszkiewicz-99,czart-11,czart-14,czart-19,Sikkema-95a,Sikkema-95,Bossche-96,Bouzerar-97,Japaridze-97,Japaridze-00,Roy-97}. 
The main efforts focused on the ground state (i.e., at $T=0$) properties of the model in one
dimension ($d=1$) at half-filling ($n=1$) 
\cite{Sikkema-95a,Sikkema-95,Bossche-96,Bouzerar-97,Japaridze-97,Japaridze-00}. 
In higher dimensions ($1<d\leq \infty $) and arbitrary electron concentration ($0<n<2$) the model has been analysed only for $J>0$ at $T=0$ \cite{Robaszkiewicz-99} and for $T>0$ \cite{czart-11,czart-19}. 
Some preliminary results for $J<0$ have been presented in \cite{czart-14,czart-15,czart-19,Dolcini-00} (for $T\geq 0$).

The $T=0$ phase diagram of the half--filled $d=1$ PK model obtained within the Hartree--Fock approximation (HFA)\cite{Robaszkiewicz-99} is in agreement with that derived by exact Lanczos diagonalizations \cite{Bouzerar-97,Japaridze-00}, the density-matrix renormalization group method \cite{Sikkema-95a,Sikkema-95,Bossche-96},  as well as with the continuum limit field theory approaches \cite{Japaridze-97,Japaridze-00}.

In previous works extensive studies of the superfluid properties of the PK with attractive $J$ ($J>0$) at the ground state and $T>0$ for  $d$-dimensional hypercubic lattice, particularly for $d=2$  square (SQ) lattice  and $d=3$ simple cubic (SC) lattice \cite{czart-11,czart-12,czart-19} were presented. 
In those works the effects of phase fluctuations on the s-wave superconductivity within the Kosterlitz--Thouless scenario for $d=2$ SQ lattice were analyzed.
It is found that due to the phase fluctuations  the gap to critical temperature ratio is substantially enhanced.
Moreover, a separation of the energy scales for the pair formation ($\sim k_BT_c$) and the phase coherence ($\sim k_BT_{KT}$) is indicated. 
For $J>0$, a continuous second-order transition to usual s-wave pairing state at $J=0^{+}$
with no additional transition for any possitive $J$ is found in all mentioned approaches  \cite{Robaszkiewicz-99,Sikkema-95a,Sikkema-95,Bossche-96,Bouzerar-97,Japaridze-97,Japaridze-00,czart-11}. 
The results of Ref. \cite{Robaszkiewicz-99} indicate that  such behavior remains unchanged in higher dimensions (including the exactly solvable case of $d=\infty $) and does not depend on the band filling (at least for alternating lattices).

In this work we study, for arbitrary $n$, the case of  repulsive pair hopping interaction $J<0$, which can favour the eta--type pairing. 
We extend previous preliminary works and present complete study of the phase diagrams, thermodynamic and electromagnetic properties of  the superconducting eta--phase for arbitrary $J<0$ and particle concentration ($0<n<2$). 
The calculations are performed for two $d$-dimensional hypercubic lattices, namely, for $d=2$ (the SQ lattice) and $d=3$ (the SC lattice). 
Results for the infinite-dimensional lattices ($d=\infty$) are also given (Appendix \ref{sec:appC}) and compared with those obtained  for the SQ and SC lattices.
The ground state $J$ versus $n$ phase diagrams are determined within the (broken symmetry) HFA.
On the diagrams we also plot,  using the Leggett's criterion \cite{Leggett-80}, the location of the crossover to the Bose--Einstein condensation (BEC) regime (cf. Eq. (\ref{BEC})).
The results presented are also contrasted with those obtained for the case of attractive $J$ (i.e., $J>0$). 

The eta-pairing superconductivity is found to be stable against the orbital (diamagnetic) pair-breaking mechanism 
\cite{Mierzejewski-04}. 
External magnetic field reduces this type of pairing mostly due to the Zeeman effect. 
According to the experimental data this mechanism is responsible for closing of the pseudogap that may 
occur in the eta-phase. 
The presence of the pseudogap has been confirmed with various experimental techniques: 
NMR \cite{Williams-97,Williams-98},
intrinsic tunneling spectroscopy \cite{Krasnov-00, Krasnov-01},
angleresolved photoemission \cite{Marshall-96,Ding-96,Norman-98}, 
infrared \cite{Basov-94} and transport \cite{Tallon-95} measurements. 
The scanning tunneling microscopy (STM) allows a direct comparison of local electronic properties in tunneling characteristics with the theoretical findings for the density of states and superconducting local gap (for review see, e.g., Refs. \cite{Balatsky-06,Krzyszczak-10} and references therein). 
Thus, in correspondence with the theoretical results for the local effective gap \cite{Kapcia-15}, the 
STM spectroscopy can be useful to distinguish s-wave and eta-pairing superconductivity in real materials.
The theory of eta-pairing superconductivity has raised recently great interest also because of experiments in photoinduced systems and cold fermionic atoms \cite{Li2020,Tindal2019,Kaneko2019,Ejima2020,Mark2020,Montorsi1996}. 

We determine the superfluid characteristics such as the critical fields $H_c$, the coherence length $\xi$, the London penetration depth $\lambda $, and the Ginzburg ratio $\kappa$ as a function of $n$ and pairing strength for the SQ and SC lattices at $T=0$ and $T>0$.  
In the analysis of these characteristics we use a linear response theory  \cite{Scalapino-93,Czart-96,Czart-96a} and the electromagnetic kernel is evaluated within the HFA Random Phase Approximation (HFA-RPA) scheme.
At $T=0$, the HFA applied to the Hubbard model and its various extensions gives reliable predictions for the ordered states properties such as, e.g., energy gap,  penetration depth, chemical potential, collective excitations, within the whole interaction range \cite{Robaszkiewicz-99,Micnas-90,Czart-96,Czart-96a}. 
Moreover, for the fermionic models with intersite interactions only, as in the considered PK model, the HFA is an exact theory for any temperature in the limit of $d\rightarrow\infty$ \cite{Muller1989}. 
For $d<\infty $,  the HFA is less reliable for $T>0$  than at $T=0$, particularly for the strong coupling limit as well as for low dimensional systems, because it neglects  phase fluctuation effects and short-range correlations.

We calculate the HFA transition temperature  $T_c$ defined as temperature at which the gap parameter vanishes ($x_{\eta}\rightarrow 0$). 
It gives the estimation for the pair--breaking temperature. 
We also present the order parameter analysis as a function of $J$ and $n$ for various lattice structures, as well as the results for $T_m$, the temperature at which minimal gap between two quasiparticle bands $E_g^{min}$ (defined later in the text) vanishes.

For $J$ and $n$ beyond the weak coupling regime, the pair formation (at temperature $T_c$) and their
condensation are two independent processes  (except $d=\infty$). 
Because the pseudogap phase is a precursor of the superconductivity, 
$T_c$ can be treated as the temperature at which the Cooper pairs start to form. 
Then, at the lower temperature, these preformed pairs undergo Bose--Einstein condensation.
This hypothesis seems to be supported by observations of the vortex–like Nernst signal above the phase transition temperature \cite{Xu-00} that evolves smoothly into the analogous signal below the  superconducting phase transition \cite{Wang-01}. 
The Meissner effect does not occur in the pseudogap phase  due to strong phase fluctuations rather than the vanishing of the superfluid density. 
In Ref. \cite{Mierzejewski-04} it is shown that the repulsive $J<0$ may lead to the occurrence of local minimum in the density of states, that is characteristic feature for pseudogap phase of underdoped cuprates \cite{Mierzejewski-04}.
The temperature dependence of the gap closing field in eta-phase differs qualitatively from the usual s-wave case and 
fits the experimental data very well. 
Such a behavior of the critical field resembles the pseudogap closing  field $H_{pg}(T)$, that has been observed in Bi$_2$Sr$_2$CaCu$_2$O$_{8+y}$ \cite{Shibauchi-01}. 
These features are not present for attractive $J>0$. 
In this case  the temperature dependence of the critical field and the gap structure are similar to those occurring for the AH model.
However, this does not mean that superconducting phase (with long-range order) in this group of compounds is eta-pairing phase.

In this paper we also investigate the effects of the Fock term on the ground state phase boundaries between the eta and the normal phases and on the eta-pairing order parameter $x_\eta$ for both SQ and SC lattices.  
Furthermore, for SC lattice we study the Fock parameter $p(T_c)$ and band narrowing at $T_c$ for the eta--pairing.

Going beyond the HFA for $d=2$ SQ lattice, the effects of phase fluctuations on the eta-pairing superconductivity are investigated.
We have compared temperature $T_c$ with a critical temperature estimated by the Kosterlitz--Thouless (K--T) scenario ($T_{KT}$).
The K--T scenario describes the  phase transition in terms of vortex pair unbinding transition, and $T_{KT}$ is determined by the universal K--T relation describing the jump of the superfluid stiffness (helicity modulus) 
$\rho _s$ at the critical temperature \cite{Kosterlitz-73,Mic-02,Denteneer-93,Denteneer-93a,Singer-98,Singer-98a}. 
Obviously  $T_{KT}$ is lower than $T_c$ due to phase fluctuation effects. 
The K--T approach was successfully applied to the $d=2$ AH model \cite{Denteneer-93,Denteneer-93a,Singer-98,Singer-98a},  
and for the models with intersite attractive interactions \cite{Chattopadhyay-97,Chattopadhyay-97a,Micnas-99,Tobijaszewska-99,czart-11},
and, in the former case, a correct behavior of $T_{KT}$ vs. the onsite attraction is found, which is in agreement with the available Quantum Monte Carlo (QMC) results \cite{Singer-98,Singer-98a}.

The paper is organized in the following way.
In Sec. \ref{sec:genfor}, the formalism and basic equations for electron thermal averages, the free energy and the chemical potential evaluated for the superconducting eta--phase are presented. 
We present also the electromagnetic kernels and the equations determining basic superfluid characteristics of the system. 
Sec. \ref{sec:resdis} is devoted to discussion of the results of numerical analysis of these equations.
In Sec. \ref{sec:resdis-GS} we present the ground state phase diagrams and analyze superconducting properties of the eta--phase (for SQ and SC lattices) as a function of the electron filling and the coupling strength.
In Sec. \ref{sec:resdis-FinTemp} we discuss the phase diagrams and superfluid characteristics at finite temperatures.
The work is concluded in Sec. \ref{sec:conclusions}. 
In the first two appendixes we summarize the analytic results for the ground state characteristics and the critical temperatures, derived in the limiting cases of strong and weak coupling  (Appendix \ref{sec:appA}) and selected results obtained for $d= \infty$ (at the ground state and $T>0$) for comparison (Appendix \ref{sec:appB}).
Appendix \ref{sec:appC} contains formulas of the densities of states used in the numerical calculations.

\section{General formulations}
\label{sec:genfor}

The current operator is derived by differentiation of the Hamiltonian (\ref{H}) with respect to the vector potential $\stackrel{\rightarrow}{A}$. 
In the standard linear approximation, the current operator is obtained as a sum of the diamagnetic and the paramagnetic parts, namely:
\begin{eqnarray}
\hat{j}_{\alpha}^{dia}(i)+\hat{j}_{\alpha}^{para}(i)  = \qquad \qquad \qquad \qquad \qquad \qquad \qquad \qquad  \\  \label{j}
\frac{e^{2} A_\alpha (i)}{\hbar ^2c}%
\left[ 
\sum_{\sigma}{ t \left(\hat{c}_{i\sigma }^{\dag}\hat{c}_{i+\vec{a}_{\alpha} \sigma} + h.c. \right)}
+ 4J\left(\hat{\rho}_i^{+}\hat{\rho}_{i+\vec{a}_{\alpha} }^{-}+h.c.\right)\right]  
\nonumber \\
+ \frac{ie}{\hbar}  \left[ 
\sum_{\sigma}  {t \left(\hat{c}_{i\sigma }^{\dag}\hat{c}_{i+\vec{a}_{\alpha} \sigma } - h.c.\right) }+
2J \left(\hat{\rho}_i^{+}\hat{\rho}_{i+\vec{a}_{\alpha} }^{-}-h.c.\right)\right], \nonumber
\end{eqnarray}
where $\alpha =x,y,z$, $\hat{\rho}_i^{+}= \hat{c}_{i\uparrow}^{\dag} \hat{c}_{i\downarrow}^{\dag}$, 
and $\hat{\rho}_i^{-}= \hat{c}_{i\downarrow} \hat{c}_{i\uparrow}$.

From the linear response theory \cite{Scalapino-93,Czart-96,Czart-96a},  
the Fourier transform of the {\em total current operator} (its expectation value) for a weak potential is derived as
\begin{eqnarray}
J_\alpha (\vec{q},\omega )=
\frac{Nc}{4\pi }\sum_{\alpha^{\prime }}
\left[
\delta_{\alpha \alpha ^{\prime }}
K_{\alpha}^{ dia }
+K_{\alpha \alpha ^{\prime }}^{ para}(\vec{q}{,\omega 
})
\right] 
A_{\alpha ^{\prime }}(\vec{q}{,\omega }). \qquad
\end{eqnarray}
The paramagnetic part is this current operator is expressed by the retarded  Green's
function (current--current) as
\begin{eqnarray}
K_{\alpha \alpha ^{\prime }}^{para}(\vec{q},\omega ) = \qquad \qquad \qquad \qquad \qquad \qquad \qquad \qquad  \label{Kpara} \\
\frac{4\pi }{c^{2}}\frac{i}{N} 
\int_{-\infty }^\infty dte^{i\omega t}\Theta(t)\left \langle
\left[\hat{j}_\alpha^{para}(\vec{q},t),\hat{j}_{\alpha ^{\prime }}^{para}(-\vec{q},0)\right]\right\rangle, 
\nonumber
\end{eqnarray}
where  $\Theta(t)$ is the Heaviside step function and $\hat{j}_\alpha^{para}(\vec{q},t)$ is the space-Fourier transform of the
paramagnetic part of the current operator (\ref{j}) in the Heisenberg
representation.

The calculations were performed for alternating lattices 
(where $\epsilon_{\vec{k}+\vec{Q}} = -\epsilon_{\vec{k}}$). 
Within HFA the free energy of the eta--phase $F_{\eta} $ is derived as:
\begin{eqnarray}
\frac{F_\eta }{N} &=&
\mu \left( n-1\right) +\frac{4}{z}Jp^{2} 
-J_{0} x_{\eta}^{2} \label{Feta} \\
&-&
\frac{1}{\beta N}\sum_{\vec{k},r}\ln \left[ 2\cosh \left( 
\frac{\beta}{2} E_{\vec{k}}^{r} \right) \right] ,
\nonumber
\end{eqnarray}
where 
\begin{equation}
E_{\vec{k}}^{r}= \epsilon_{\vec{k}} + r A_{\vec{k}} , 
\label{Ekr3}
\end{equation}
and 
$A_{\vec{k}}=\sqrt{\mu^{2} + J_{0}^{2} x_{\eta}^{2}}$, $r=\pm 1$. 
For hypercubic lattices with NN hopping: $\epsilon_{\vec{k}}=-\widetilde{t\;}%
\gamma_{\vec{k}}$, $\widetilde{t}=t+2pJ/z$, $\gamma_{\vec{k}}=2 \sum_{\alpha} \cos
k_{\alpha}$, $\alpha =x,y,\ldots$, $J_{0}=zJ$, $z$ is the number of NN, and $\beta
=1/(k_{B}T)$.
The sum $\sum_{\vec{k}}$ (here and in all further places in the work) denotes the summation over all vectors $\vec{k}$ in the first Brillouin zone.

Two branches $E_{\vec{k}}^{+}$ and $E_{\vec{k}}^{-}$ exist in the electronic spectrum of the eta--phase  and the minimal gap between the lower and higher band (defined as $E_{g}^{min}=\mbox{min}E_{\vec{k}}^{+} - \mbox{max}E_{\vec{k}}^{-}$) can be
either positive or negative (depending on model parameters and temperature). 

The eta--pairing order parameter 
$x_{\eta} =(1/N)\sum_{i}\exp (\vec{Q} _i\cdot \vec{R}_i) \langle \hat{c}_{i\downarrow }\hat{c}_{i \uparrow
}\rangle $,
the chemical potential $\mu $, and 
the Fock term $p=1/(4 N) \sum_{\vec{k},\sigma }\gamma_{\vec{k}} \langle \hat{c}_{\vec{k}\sigma }^{\dag}\hat{c}_{\vec{k}\sigma }^{}\rangle $
are determined by the equations 
\begin{equation}
\frac{\partial F_\eta }{\partial x_\eta } =0 , \qquad 
\frac{\partial F_\eta }{\partial \mu }=0 , \qquad
\frac{\partial F_\eta }{\partial p}=0 , 
\end{equation}
and their explicit forms are the following:
\begin{eqnarray}
\label{eqVeta} \frac{1}{J_0} & = & \frac{-1}{4N} \sum_{\vec{k},r=\pm 1} \frac{r}{A_{\vec{k}}} \tanh \left( \frac{\beta}{2} E_{\vec{k}}^{r}\right),   \\
\label{eqneta}
n-1 & = & \frac{1}{2N} \sum_{\vec{k},r=\pm 1} r \frac{\mu}{A_{\vec{k}}} 
 \tanh \left( \frac{\beta}{2} E_{\vec{k}}^{r}\right),
\\
\label{eqpeta}
p & = & \frac{-1}{8N} \sum_{\vec{k},r=\pm 1} \gamma_{\vec{k}} \tanh \left( \frac{\beta}{2}
E_{\vec{k}}^{r}\right).  
\end{eqnarray}
The expressions for  $F$, $\mu$, and $p$ in the normal (N) phase are derived by taking $x_{\eta}=0$ in Eqs. 
(\ref{Feta}),  (\ref{eqneta}), and (\ref{eqpeta}), respectively.

The {\em magnetic penetration depth} can by determined as the transverse part of the total kernel 
in the static limit (the local approximation, i.e., the London limit) by 
\begin{equation}
\lambda (T)=\left[-K_{x}^{dia}-K_{xx}^{para}(\omega =0)\right]^{-1/2}.  \label{lambda_xx}
\end{equation}
The expression for $K_x^{dia}$ is:
\begin{eqnarray}
K_{\alpha}^{dia} &=& K_{t}+K_{J}, \label{K-alpha-dia} \\
K_{t} &=&\frac{8\pi e^{2}}{\hbar^{2}c^{2}a}\frac{1}{N} |t|\sum_{\vec{k}\sigma } \langle \hat{c}_{\vec{k}\sigma }^{\dag}\hat{c}_{\vec{k}\sigma}^{\ } \rangle \cos(k_{\alpha}),  
\\
K_J &=&\frac{32\pi e^{2}}{\hbar^{2}c^{2}a}\frac{1}{zN}\langle -\sum_{\vec{k}} J_{\vec{k}}
\hat{\rho}_{\vec{k}}^{+}\hat{\rho}_{\vec{k}}^{-}\rangle ,
\end{eqnarray}
where $a$ is the lattice constant.  
$J_{\vec{k} } $ are the space--Fourier transforms of  $J_{ij}$.
The explicit expressions for $K_t$ and $K_J$ obtained within presented
approximation scheme take the forms:
\begin{eqnarray}
K_{t} &=& -\frac{8\pi e^{2}}{\hbar^{2}c^{2}a}\frac{1}{2N}|t|\sum_{\vec{k},r=\pm
1} \cos \left( k_{\alpha} \right) \tanh \left( \frac{\beta}{2} E_{\vec{k}}^{r}\right) \nonumber \\
& = &  \frac{8\pi
e^{2}}{\hbar^{2}c^{2}a}\frac{\langle E_{kin}\rangle}{z} ,  \label{K_dia} \\
K_{J} &=& - \frac{32\pi e^2}{\hbar^{2}c^{2}a} \frac{|J_{0}|(x_{\eta}^{ })^{2}}{z}  ,
\end{eqnarray}
where $x_{\eta}^{ }$, $p$ and $\mu $ are determined by Eqs. (\ref{eqVeta})--(\ref{eqpeta}), and 
$\langle E_{kin} \rangle$ is the average value of the kinetic energy.

In the ground state the paramagnetic part $K^{para}_{\alpha\alpha'}$ of the kernel may be significant in determining 
$\lambda$ only for nonlocal (Pippard) superconductors when the correlation length is greater than the penetration depth $\lambda $. 
The short-coherence length superconductors, which are studied in this work, are in the opposite limit, i.e., the London limit. 
In such a case the  penetration depth at $T=0$ is determined entirely by the $q\rightarrow 0$ limit
of the kernel.
In this limit  the$K^{para}_{\alpha\alpha'}$ vanishes and $\lambda $ is given by 
\begin{equation}
\lambda =\frac{1}{\sqrt{-K^{dia}}}.
\label{lambda}
\end{equation}
Using the values of $\lambda$ [Eqs. (\ref{lambda}) and (\ref{lambda_xx})], and the difference of the free energy between the normal (N) and eta--phases, one determines the thermodynamic critical field $H_c$ as well as the Ginzburg-Landau correlation length $\xi_{GL}$ as
\begin{eqnarray}
\frac{H_{c}^{2}(T)}{8\pi } &=&\frac{F_{N}(T)-F_{\eta}(T)}{Na^{3}},  \nonumber \\
\xi_{GL} &=&\frac{\Phi_{0}}{2\pi \sqrt{2}\lambda H_{c}},  \label{Hcxi} 
\end{eqnarray}
where $\Phi_{0}=hc/(2e)$.
One can also obtain the estimates of two critical fields $H_{c1}\simeq H_{c} \ln(\kappa)/ \kappa $ and $H_{c2}=\Phi _{0}/(2\pi \xi_{GL}^{2})$, where $\kappa =\lambda /\xi_{GL}$.

From Eqs. (\ref{eqVeta})--(\ref{eqpeta}) one can calculate the HFA transition temperature $T_c$ (defined as temperature at which the gap amplitude vanishes, $x_{\eta} \rightarrow 0$). 
It is an estimation of the eta-pair-formation temperature.

\begin{figure*}[t]
\centering
\includegraphics[width=\sizetwo]{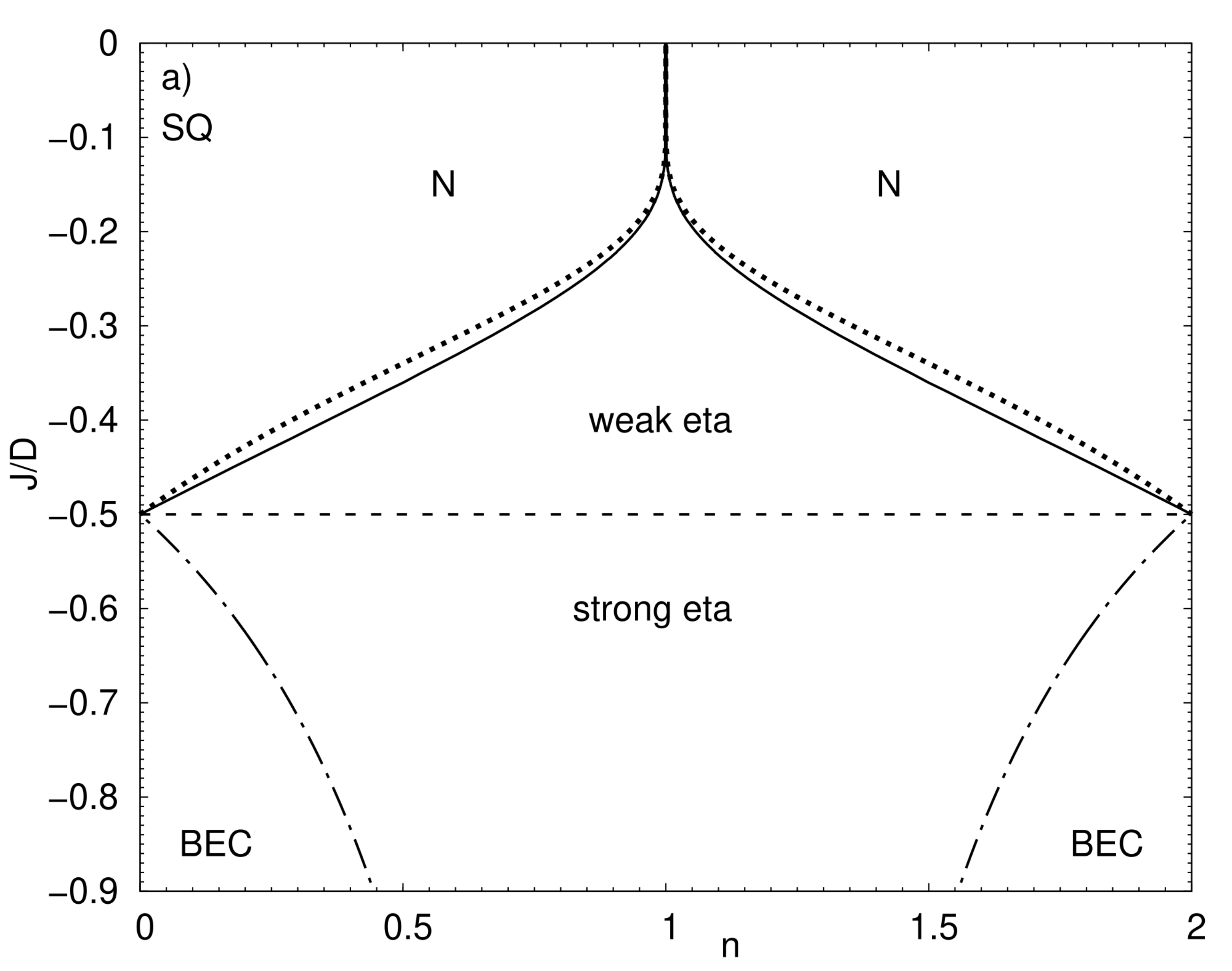}
\includegraphics[width=\sizetwo]{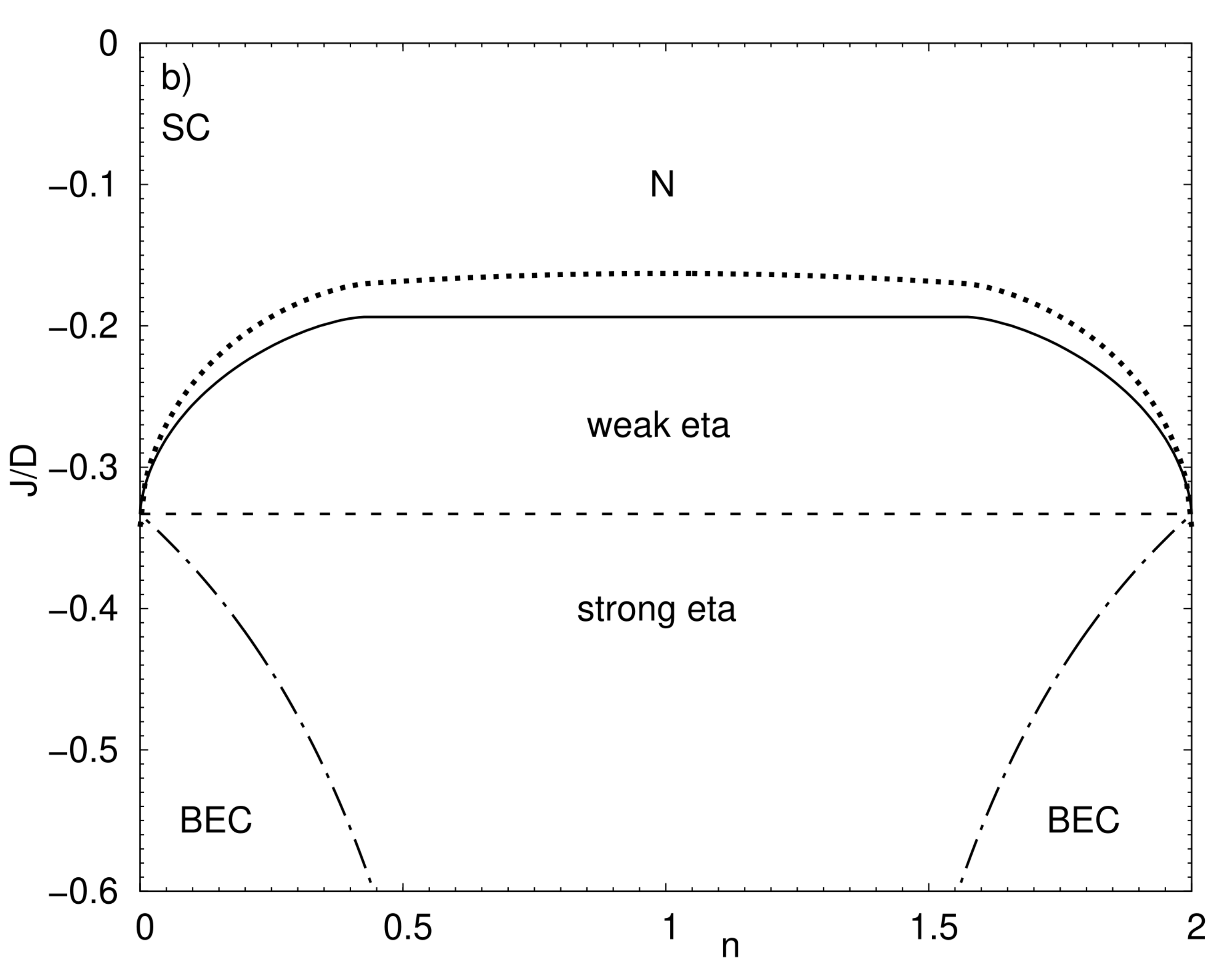}
\caption{The ground state boundaries between the eta and normal (N) states for (a) $d=2$ SQ and (b) $d=3$ SC lattices ($D=2dt$) as a function of electron concentration $n$.   
Dotted and solid lines denote the phase boundaries for the PK model with and without, respectively, the Fock term.
\emph{Strong}--eta and \emph{weak}--eta regimes are separated with the horizontal dashed line (defined in the text), whereas dashed-dotted lines denote the location of the crossover to BEC regime.
}\label{fig:diag-U0-vsn}
\end{figure*}

Except for the case of infinite dimensions (i.e., $d \rightarrow \infty $), the superconducting phase transition will occur at the temperature, which is lower than $T_c$ because of the phase fluctuation effects.  
As we pointed out in Sec. \ref{sec:intro}, for d $=2$ lattice the temperature of superconducting transition can be derived within the Kosterlitz--Thouless theory  \cite{Kosterlitz-73,Mic-02,Denteneer-93,Denteneer-93a,Singer-98,Singer-98a}, which describes the transition in terms of vortex pair unbinding transition.
One can find this transition at $T_{KT}$ by using the K--T relation for the universal jump of the superfluid stiffness $\rho_s $. 
The critical temperature $T_{KT}$ for the K--T transition is obtained by calculating the $\rho_s$ as a function of temperature and then by comparison with the K--T relation between $T_{KT}$ and $\rho _s$: 
\begin{equation}
k_{B} T_{KT}= Q \rho _s(T_{KT}),  \label{kBTc} 
\end{equation}
where $Q \simeq 0.898$ (Monte-Carlo estimates for $d=2$ XY model \cite{czart-11,Gupta-88}) and 
$\rho_{s}$ is the superfluid stiffness, which is related to the London penetration depth [Eqs. (\ref{lambda_xx}), (\ref{K-alpha-dia}), and (\ref{Kpara})]: 
\begin{eqnarray}
\rho_{s}(T) &=& \frac{\hbar^{2}c^{2}a}{16\pi e^{2}}\lambda ^{-2} \nonumber \\
&=& \frac{-\hbar^{2}c^{2}a}{
16\pi e^2} \left[ K_x^{dia}+K_{xx}^{para}(\omega =0)\right] .
  \label{gs}
\end{eqnarray}
Thus, $T_{KT}$ is determined as a solution of four self-consistent equations (\ref{eqVeta})--(\ref{eqpeta}) and (\ref{kBTc}).
An upper bound for the $T_{KT}$ can be obtained by
\begin{eqnarray}
k_B\tilde{T}_{KT} = Q \rho_s(T=0) =  Q \frac{\hbar^2c^2a}{16\pi e^2} \lambda^{-2}(0)  ,
\label{TKTtylda}
\end{eqnarray}
where $\lambda$ is given by Eq. (\ref{lambda}).

Let us underline that $T_{KT}$ determined from these equations gives only an upper bound of the actual K--T transition temperature, because the HFA expression (\ref{gs}) does not include  renormalization of $\rho_s$ caused by
topological excitations (vortex-antivortex pairs) \cite{Denteneer-93,Denteneer-93a,Singer-98,Singer-98a}. 
Moreover, $\rho_s$ is approximated by its ground state value.

\section{Results and discussion}
\label{sec:resdis}

In this section a comprehensive analysis of the thermodynamic and electromagnetic properties of the eta--phase of the model (\ref{H}). 
The calculations were carried out for $d=2$ SQ, $d=3$ SC, and $d=\infty$ (Appendix \ref{sec:appB}) lattice structures, for arbitrary electron concentration $n$ $(0<n<2)$ and repulsive interactions $J<0$, both at $T=0$ and $T>0$.
We have also investigated for SQ and SC lattices the crossover to the Bose--Einstein condensation (BEC) regime. 
At $T=0$, the crossover can be located using the Leggett's criterion, which defines the BEC crossover point from the condition that the chemical potential in the superconducting phase reaches the bottom of the
electronic band \cite{Leggett-80,Nozieres-85}, i.e., from 
\begin{equation}
\mu_{\eta}=-B/2,
\label{BEC}
\end{equation}
where $B=2zt$ is the band-width and  $\mu_\eta$ is determined from the self--consistent equations (\ref{eqVeta})--(\ref{eqpeta}) solved at $T=0$. 


Note that, because Hamiltonian (\ref{H}) exhibits the electron-hole symmetry (in the considered here the nearest neighbours case), all the  presented plots are symmetric under the transformation $n \rightarrow 2-n$ (i.e., with respect to half-filling $n=1$).

\subsection{The ground state of the model}
\label{sec:resdis-GS}

In the following, we present  the ground state phase diagrams of the model and the analysis of the 
superfluid characteristics at $T=0$ as a function of concentration $n$ and repulsive $J$. 

\begin{figure*}[t]
\centering
\includegraphics[width=\sizetwo]{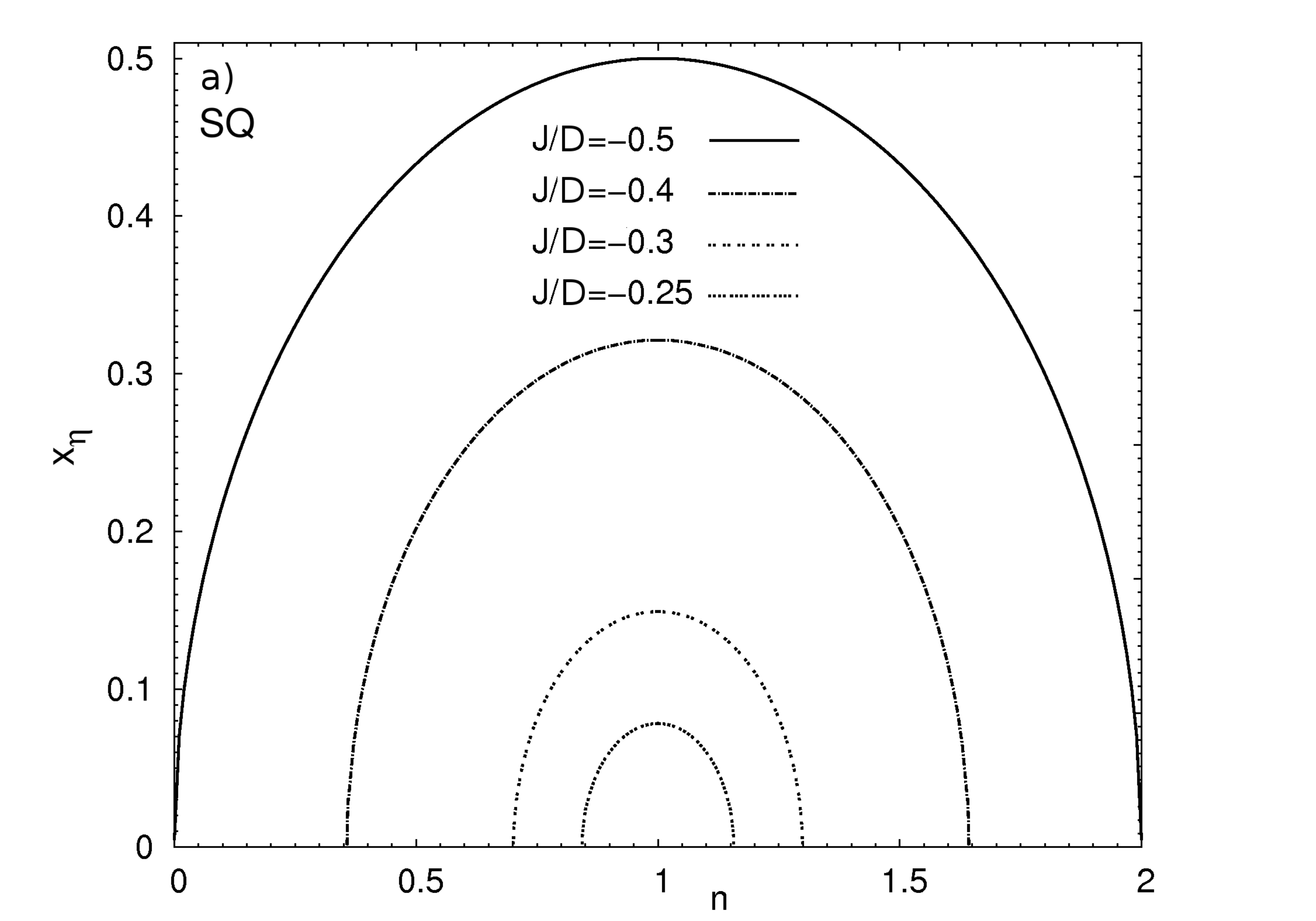}
\includegraphics[width=\sizetwo]{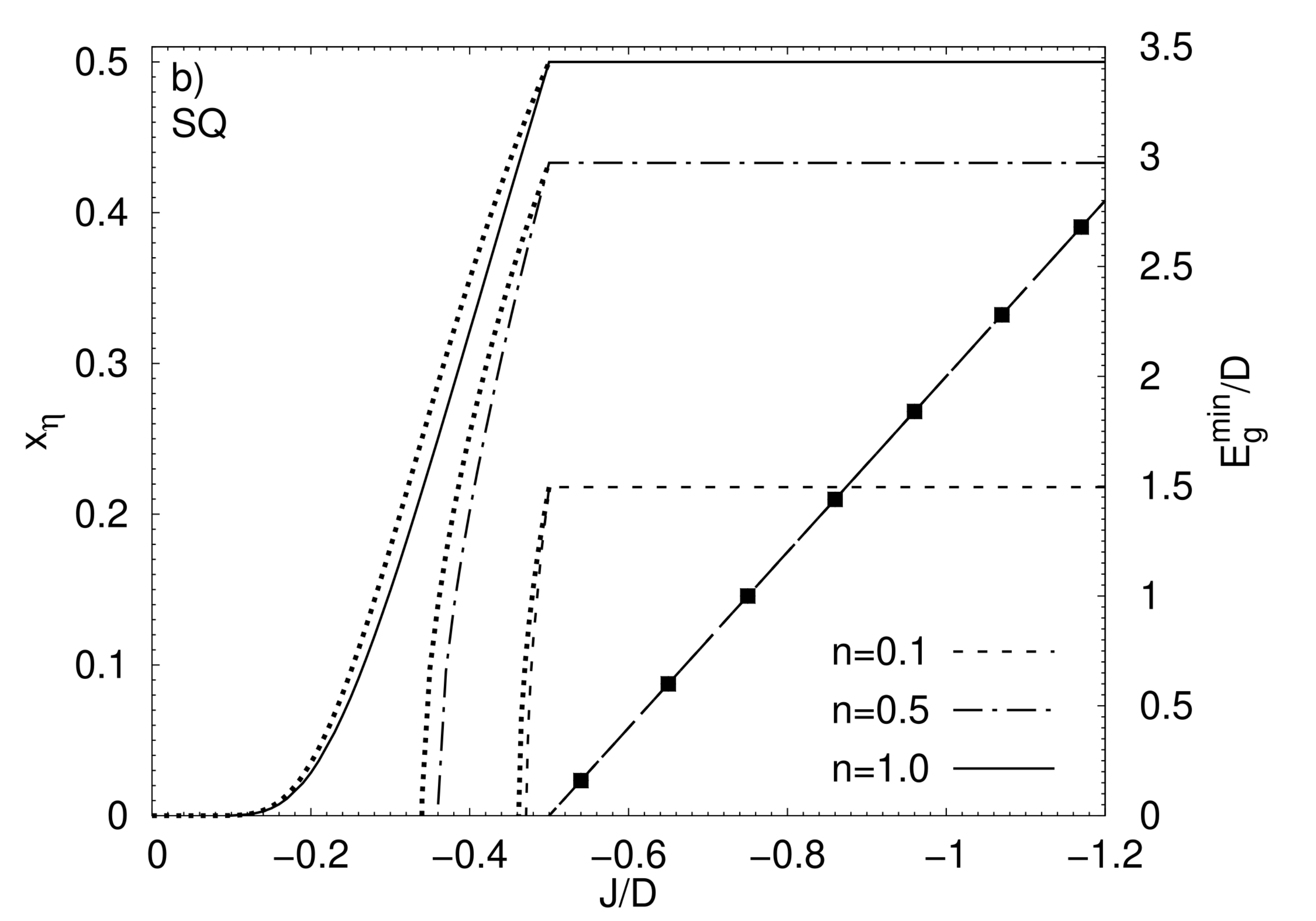}
\caption{The ground state plots of (a) the eta--pairing order parameter $x_\eta$ as a function of concentration $n$ for several fixed values of $J/D$:  $J/D=-0.25$, $J/D=-0.3$, $J/D=-0.4$, $J/D=-0.5$; and (b) the eta--pairing order parameter $x_\eta$ and the gap between the lower and higher quasiparticle band $E_g^{min}$ as function of interaction $J$  for several fixed values of $n$: $n=1.0$, $n=0.5$, $n=0.1$. 
The $E_g^{min}$ versus $J$ lines are denoted with black squares ($\blacksquare$, they are the same for all $n$). 
For $n=1$ critical point is at $J_c=0$.
The lines for the case of the PK model without the Fock term are plotted with line styles defined in the plot legend and respective effects of the Fock term are plotted as dotted lines.
Results for the SQ lattice ($D=4t$).
}\label{fig:x-Eg-vsn}
\end{figure*}

\begin{figure*}[t]
\centering
\includegraphics[width=\sizetwo]{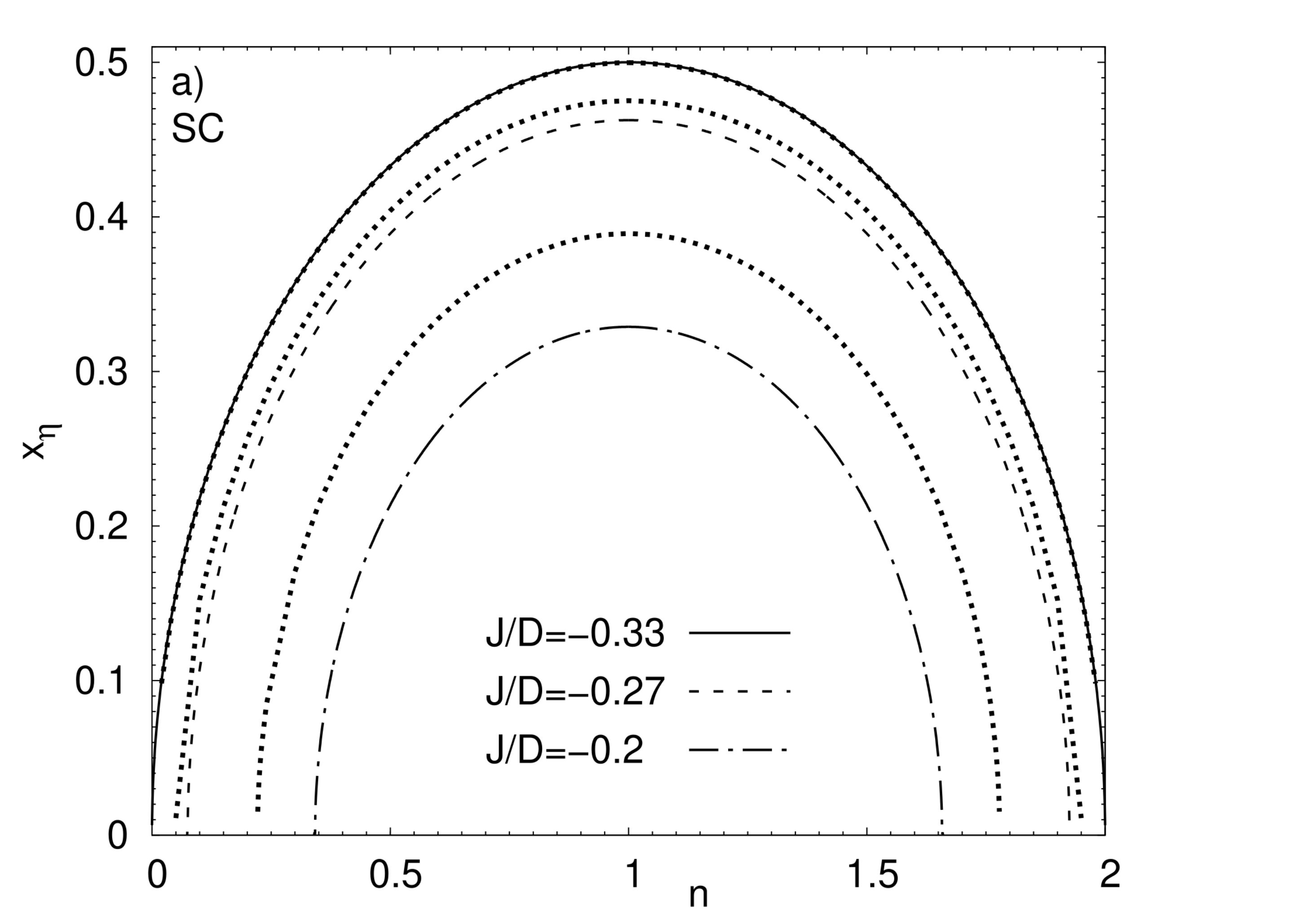}
\includegraphics[width=\sizetwo]{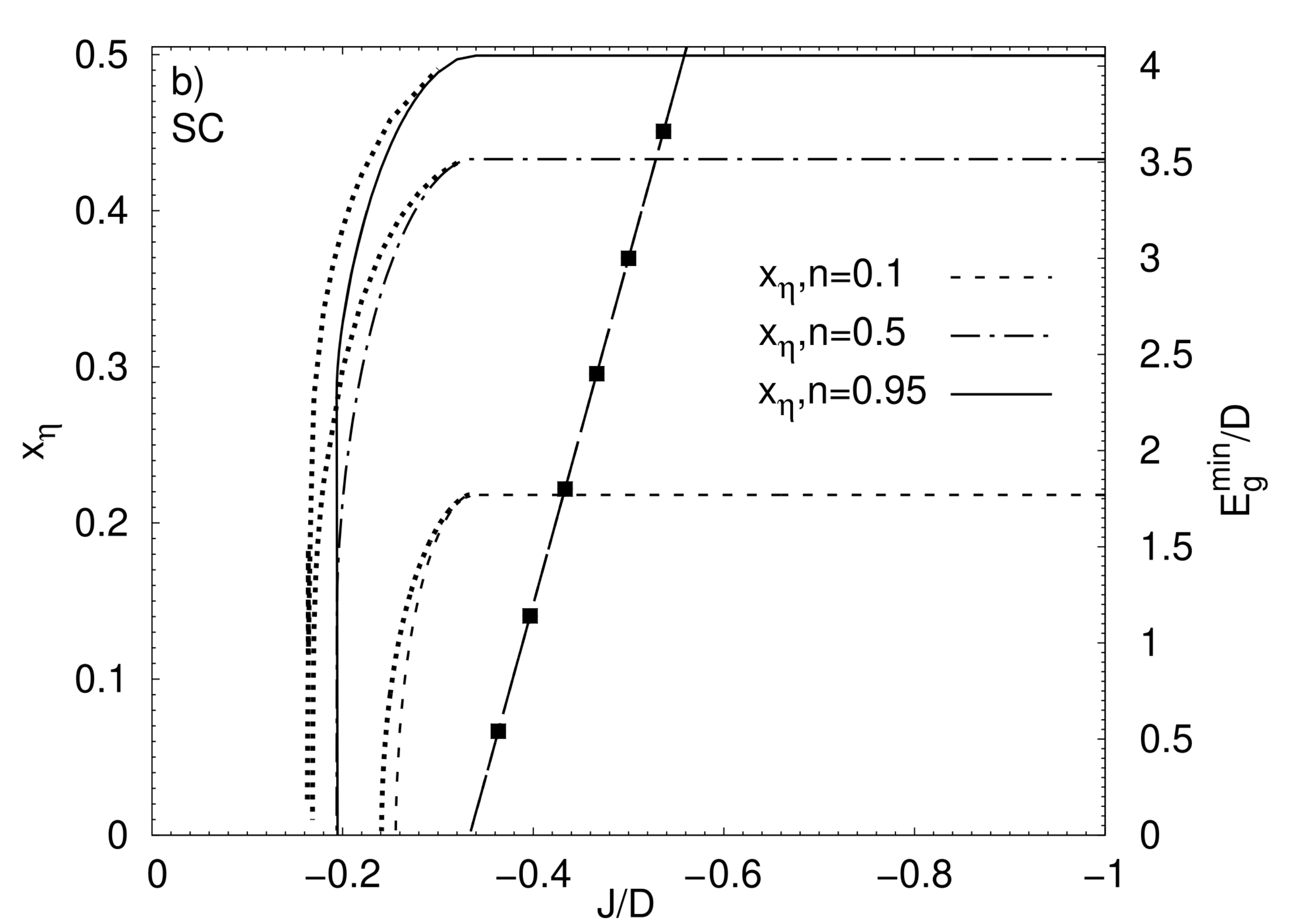}
\caption{The ground state plots of the eta--pairing order parameter $x_\eta$  as a function of (a) concentration $n$ for several fixed values of $J/D$:  $J/D=-0.2$, $J/D=-0.27$, $J/D=-0.33$; and (b) the eta--pairing order parameter $x_\eta$ and the gap between the lower and higher quasiparticle band $E_g^{min}$ as function of interaction $J$  
for several fixed values of $n$:  $n=0.1$, $n=0.5$, $n=0.95$.  
The $E_g^{min}$ versus $J$ lines are denoted with black squares ($\blacksquare$, they are the same for all $n$). 
The lines for the case of the PK model without the Fock term are plotted with a line styles 
defined in the plot legend and respective effects of the Fock term are plotted as dotted lines.
Results for the SC lattice ($D=6t$).
}\label{fig:x-Eg-vsn-sc}
\end{figure*}

\subsubsection{The phase diagram and order parameter at $T=0$}

The transition into the eta--pairing superconducting phase [favored by the repulsive $J$ ($J<0$)], is found
to occur at $T=0$ above some critical value $|J_c|$, see Fig. \ref{fig:diag-U0-vsn} (the SQ and SC lattices) and Fig. \ref{fig:diag-U0-vsn-inf} for $d=\infty$ (cf. also preliminary results in \cite{czart-14,czart-15,czart-19}).
The critical value $J_c$ depends on the lattice structure, i.e., on the form of the density 
of states $D(\epsilon )$, as well as on the band filling $n$. 
In contrast to the s-wave case, the eta--pairing phase do never exhibit standard BCS-like features.
As we find in Figs. \ref{fig:diag-U0-vsn} and \ref{fig:diag-U0-vsn-inf}, with increasing $|1-n|$,  the $T=0$ phase boundaries between the eta--pairing and N state  are shifted towards higher values of $|J|$.
Thus, in a certain range of interaction parameter $J$ the transition from the superconducting eta--phase to the N state can be realized by changing the electron density.
In particular, the strongest $n$-dependence is observed for the SQ lattice, where at $n=1$ due to the van Hove singularity $J_c=0$ and  the eta-pairing phase is stable for any $J<0$. 
In contrast, for the SC lattice and $d=\infty$ structures, $|J_c|$ significantly depends on $n$ only for 
low carriers concentration and never $J_c\rightarrow 0$. 
For SC lattice, in the case of the PK model without the Fock term for $0.4 \lesssim  n \lesssim 1.6$, $J_c$ 
does not depend on $n$, wheres in the case with the Fock term, $J_c$ weakly depends on $n$, similarly  
as for $d=\infty$ lattice, in the analogous range of $n$ for the case of the PK model without the Fock term.
Notice that inclusion of the the Fock term in to the PK model equations does not change qualitatively the phase diagrams.
The Fock term reduces $|J_c|$ and the highest reduction is observed at half--filling for the SC lattice and close to the middle between the half-filled and empty (fully occupied) band limits for the SQ lattice, for which the Fock term  disappears with $n \rightarrow 1$.

For any fixed $n$, a second characteristic value of $J < 0$ (denoted as $J_{c1}$) also exists and $\left| J_{c1}\right| \geq $ $\left| J_c\right| .$
For any $d$-dimensional hypercubic lattices with NN hopping only (except for Gaussian DOS for the $d=\infty$ lattice), $\left| J_{c1}\right| =2t$ for any $n$ (see also dotted line in Figs. \ref{fig:diag-U0-vsn} for semi-elliptic DOS).

\begin{figure*}[t]
\centering
\includegraphics[width=\sizetwo]{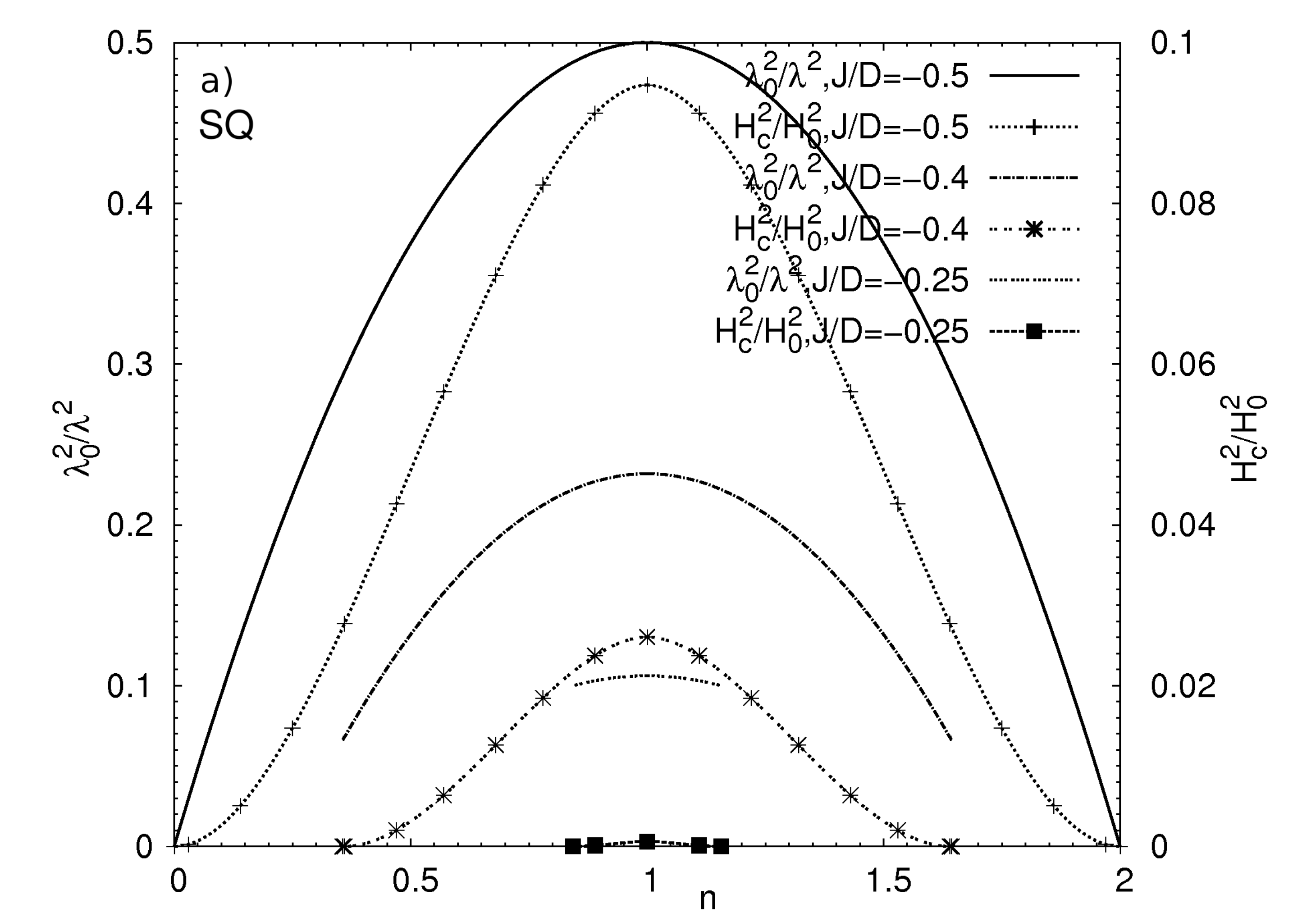}
\includegraphics[width=\sizetwo]{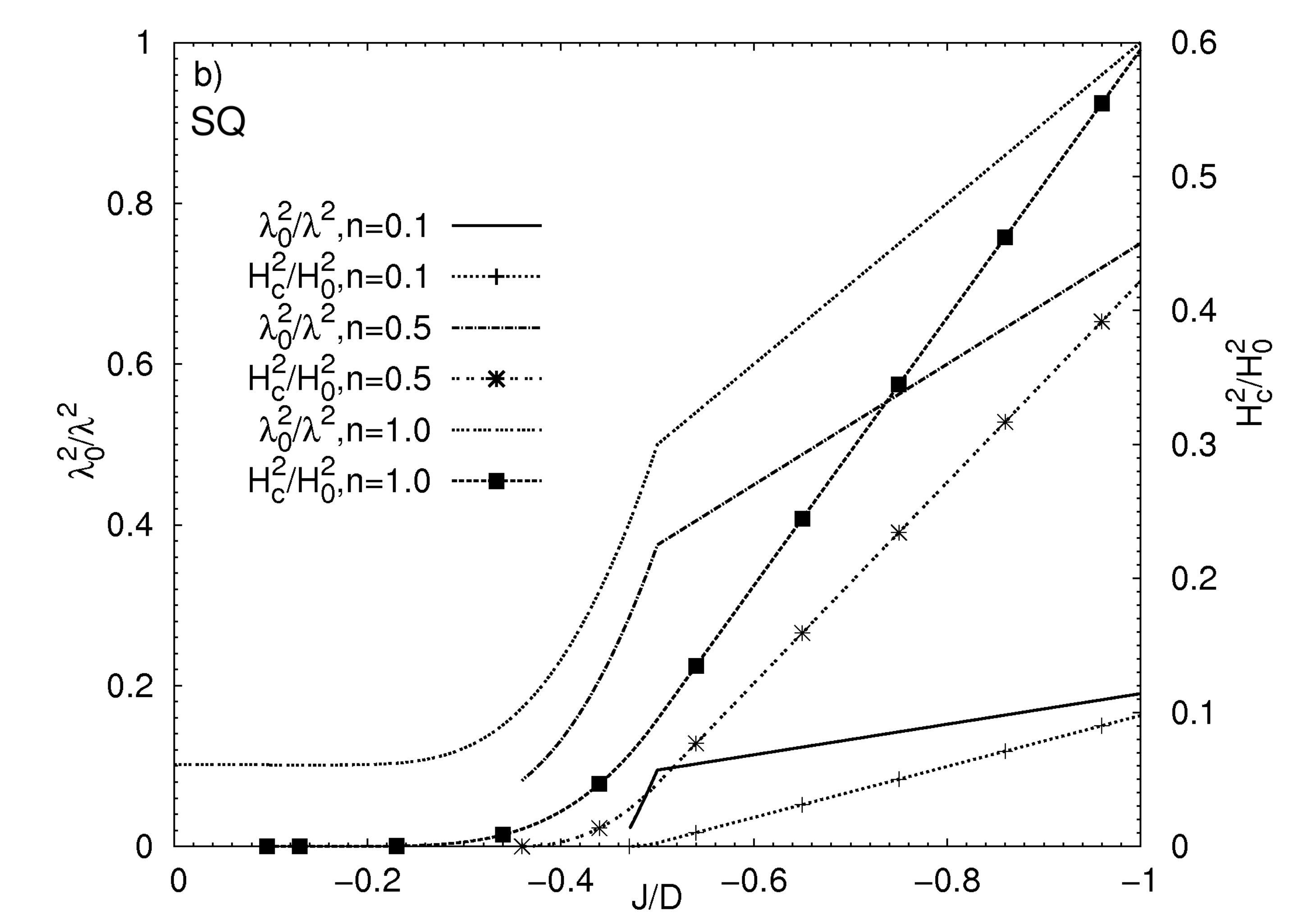}
\caption{
(a) Concentration $n$ and (b) interaction $J$ 
dependencies of the inverse square penetration depth $\lambda_0^2/\lambda^2$ 
[$\lambda_0=(\hbar c / e)\sqrt{a^{d-2}/(4\pi D)}  $], 
and the reduced square value of thermodynamic critical field 
$H_c^2/H_0^2$ $(H_0^2=4\pi D/a^d)$ at $T=0$.
Panel (a) is plotted for fixed values of $J/D$: $J/D=-0.25$; $J/D=-0.4$ and $J/D=-0.5$,
whereas panel (b) for fixed  
$n$: $n=1.0$; $n=0.5$ and $n=0.1$.
 $H_c^2/H_0^2$ -- lines with symbols, 
$\lambda_0^2/\lambda^2$ -- lines without symbols (as labeled, for SQ lattice).
}\label{fig:l-Hc-vsn}
\end{figure*}

\begin{figure*}[t]
\centering
\includegraphics[width=\sizetwo]{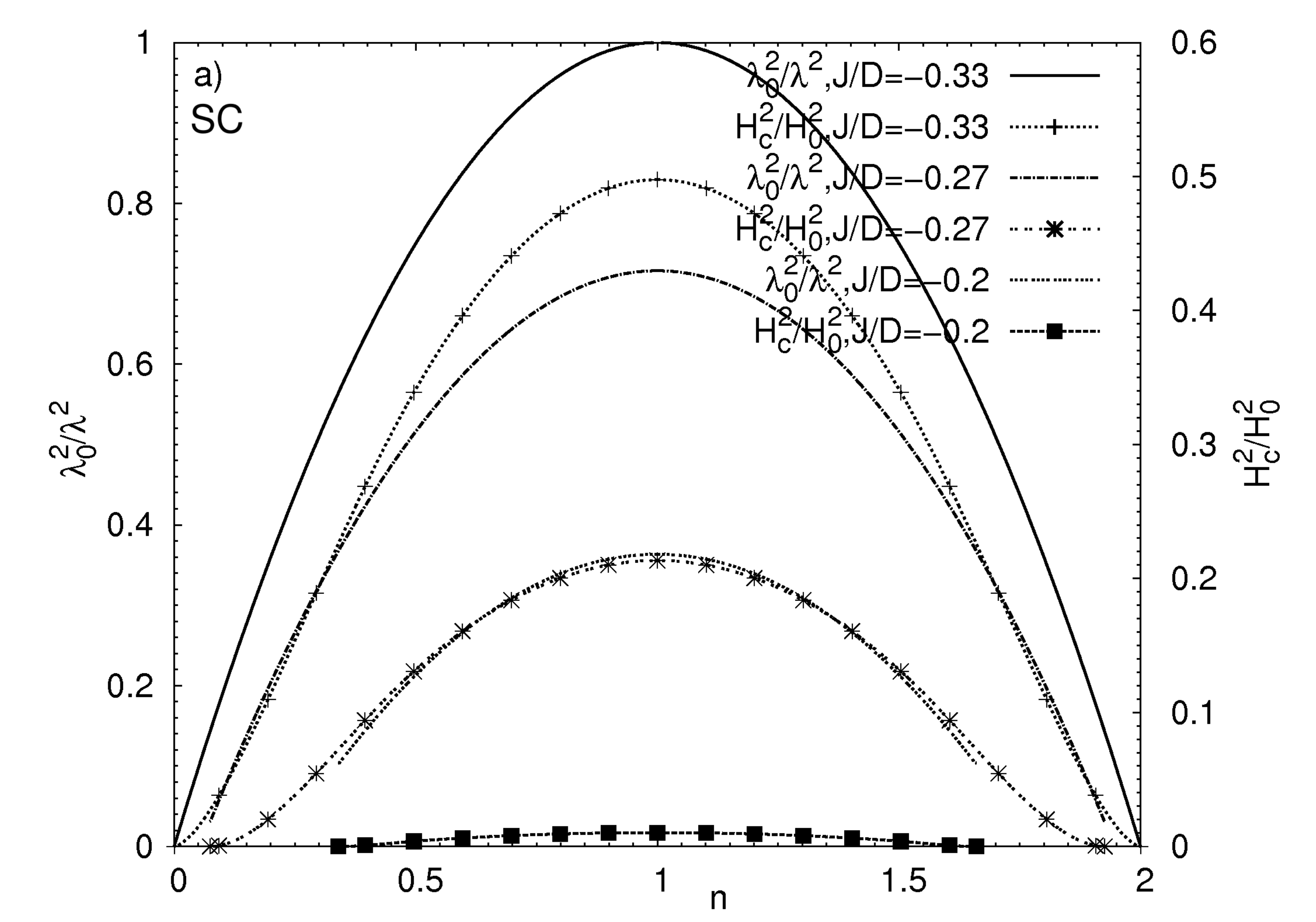}
\includegraphics[width=\sizetwo]{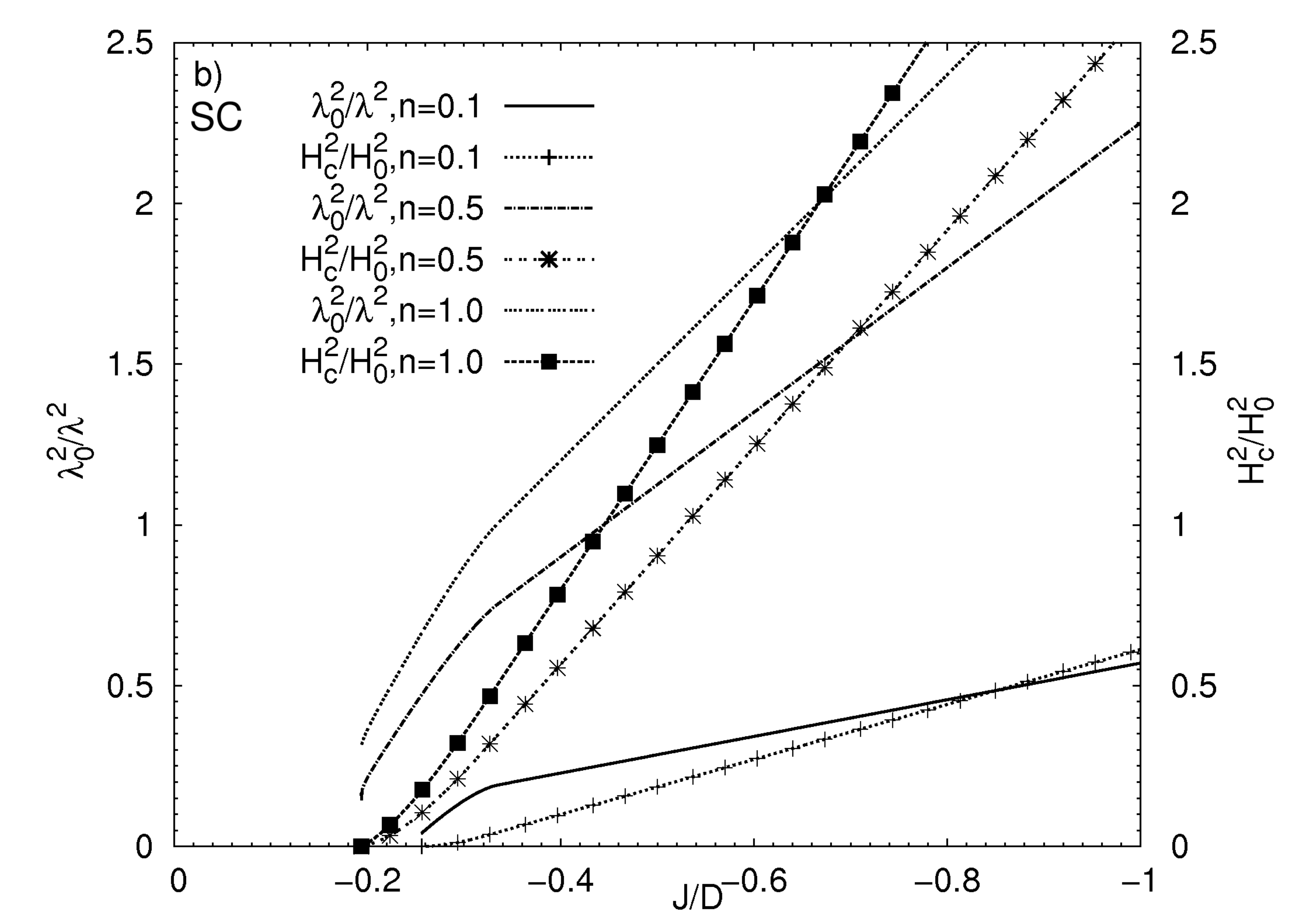}
\caption{
(a) Concentration $n$ and (b) interaction $J$ 
dependencies of the inverse square penetration depth $\lambda_0^2/\lambda^2$ 
[$\lambda_0=(\hbar c / e)\sqrt{a^{d-2}/(4\pi D)}  $], 
and the reduced square value of thermodynamic critical field 
$H_c^2/H_0^2$ $(H_0^2=4\pi D/a^d)$ at $T=0$.
Panel (a) is plotted for fixed values of $J/D$: $J/D=-0.2$; 
$J/D=-0.27$ and $J/D=-0.33$, 
whereas panel (b) for fixed  
$n$: $n=1.0$; $n=0.5$ and $n=0.1$.
 $H_c^2/H_0^2$ -- lines with symbols, 
$\lambda_0^2/\lambda^2$ -- lines without symbols (as labeled, for the SC lattice).
}\label{fig:l-Hc-vsn-sc}
\end{figure*}

For $\left| J\right| >\left| J_{c1}\right| $ the ground state of the model is characterized by a non-zero gap between the lower and higher quasiparticle band, i.e., $E_g^{min}\equiv \min E_k^{+}-\max E_k^{-} > 0$  
and by the order parameter taking its maximum value (which is the same as in the zero-bandwidth limit \cite{KRM2012,KR2013,KapciaAPPA2014}) $x_\eta ^{\max }=\frac{1}{2}\sqrt{n\left( 2-n\right) }$ (see Figs. \ref{fig:x-Eg-vsn} and \ref{fig:x-Eg-vsn-sc}). 
We define this state as the \emph{strong} eta--pairing phase (in analogy with the strong ferromagnet).  
On the other hand, for $\left| J_c\right| < \left| J\right| <\left| J_{c1}\right|$ the gap $E_g^{\min }<0$ and the order parameter $x_{\eta} <x_{\eta}^{\max }$. 
Thus, we call this state as the \emph{weak} eta--pairing phase (analogous to the weak ferromagnet). 
At $J=J_{c1}$, $E_g^{min}=0$ and the smooth crossover between the \emph{weak} and \emph{strong} eta--phase 
takes place.
For $E^{min}_g (T) < 0$ the quasiparticle DOS in the eta-pairing state is finite for arbitrary energy, but a local minimum in the DOS can appear at the Fermi level and the system can exhibit a pseudogap behavior  \cite{Mierzejewski-04}.
The \emph{weak} eta-phase is stable only within a restricted range of concentration and within this phase $E_g^{min}<0$ at any $T<T_c$.
The range shrinks with decreasing $|J|$ and the eta-phase disappears for $J\rightarrow J_c$.
For NN hopping only, the $E_g^{min}(T=0)$ does not depend on $n$.   
Thus, in the Figs. \ref{fig:x-Eg-vsn}(b), \ref{fig:x-Eg-vsn-sc}(b)  and \ref{fig:x-Eg-vsn-inf}(b) (in Appendix \ref{sec:appB}), the lines of $E_g^{min}(J)$ for different values of $n$ overlap, whereas each line for $x_{\eta}(J)$ ends (vanishes) at different critical $|J_c|$, which is dependent on $n$.

In Fig. \ref{fig:diag-U0-vsn} we also have marked the location of the crossover to  the Bose--Einstein Condensate (BEC) regime (cf. Eq. (\ref{BEC})). 
For the SQ and SC lattices at $T=0$, the values of $|J|$ at which the crossover occurs increase with decreasing $|1-n|$.
Thus, in definite range of $|J|$ the crossover to BEC can be realized by changing the electron density.

\begin{figure*}[t]
\centering
\includegraphics[width=\sizetwo]{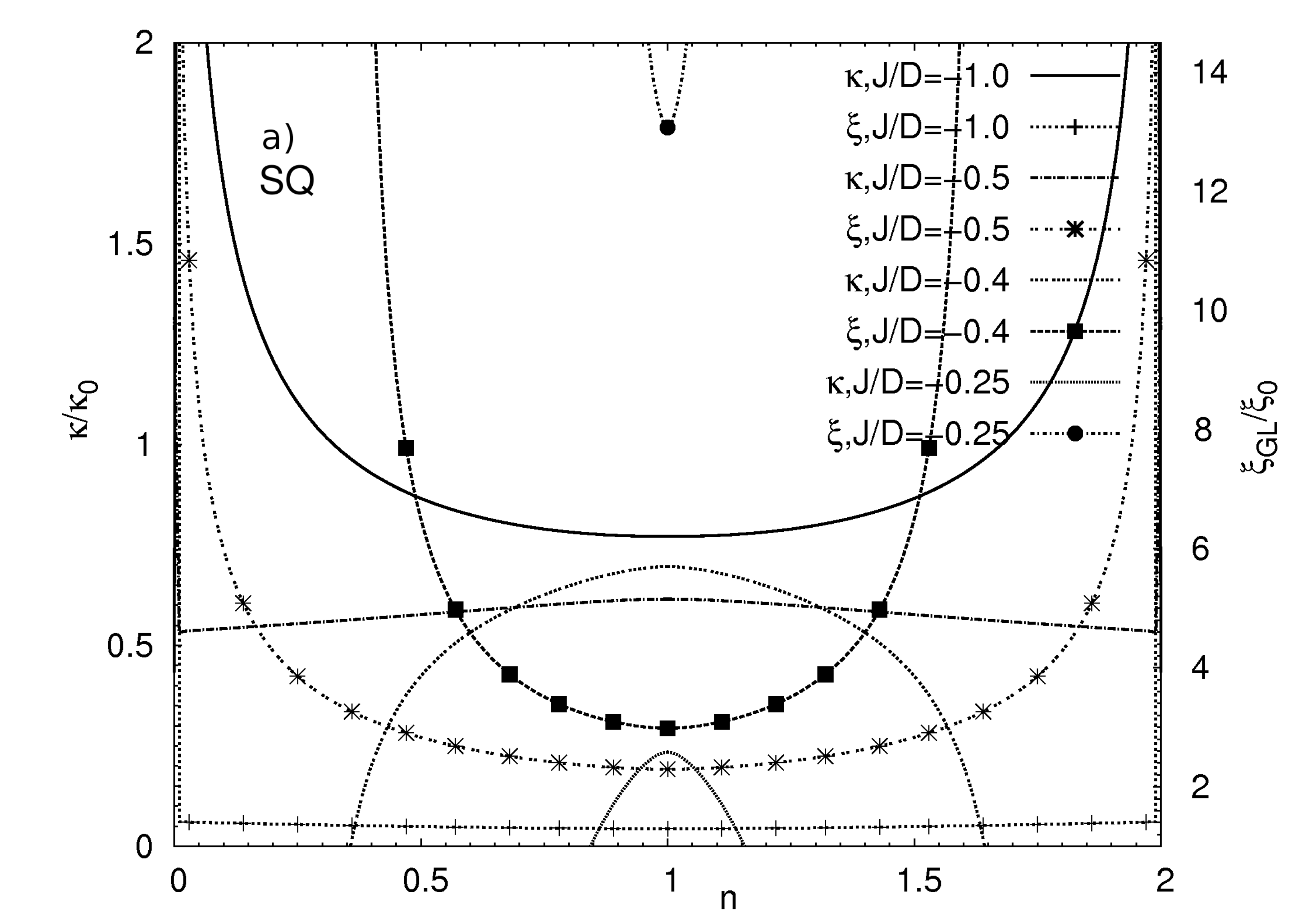}
\includegraphics[width=\sizetwo]{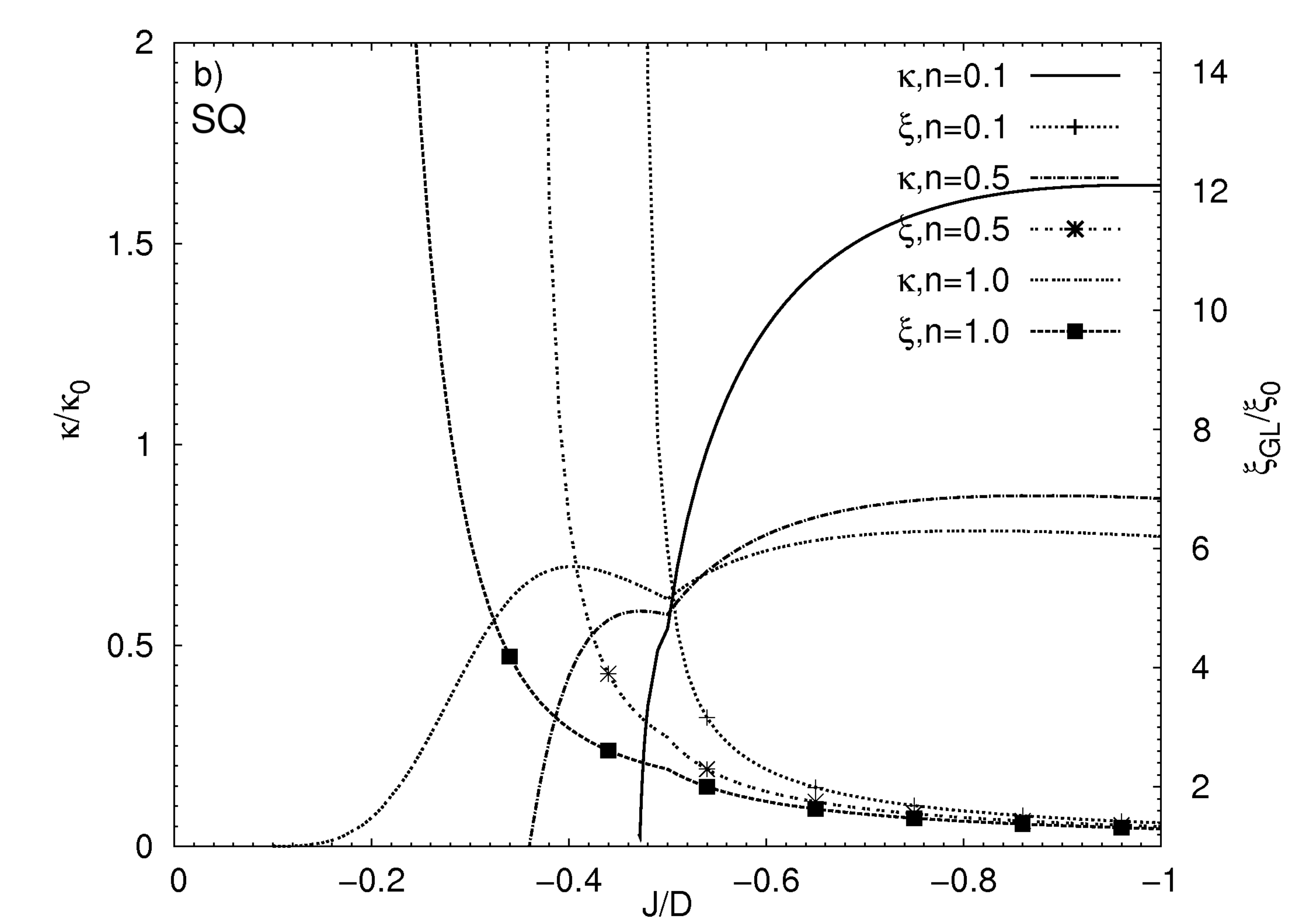}
\caption{
(a) Concentration $n$ and (b) interaction $J$  
dependences of the GL coherence 
length $\xi_{GL}/\xi _0$ ($\xi _0=a/\sqrt{2z}$, lines with symbols) and the 
Ginzburg ratio $\kappa /\kappa_0$, ($\kappa _0=(\hbar c/e)\sqrt{2}/\sqrt{\pi
Da^{4-d}}$, lines without symbols) at $T=0$.  
Panel (a) is plotted for fixed values 
of $J$: $J/D=-0.25$, $J/D=-0.4$, $J/D=-0.5$, $J/D=-1.0$, while panel (b) 
for fixed $n$: $n=0.1$, $n=0.5$, $n=1$ (for the SQ lattice).
}\label{fig:k-xi_vsn}
\end{figure*}

\begin{figure*}[t]
\centering
\includegraphics[width=\sizetwo]{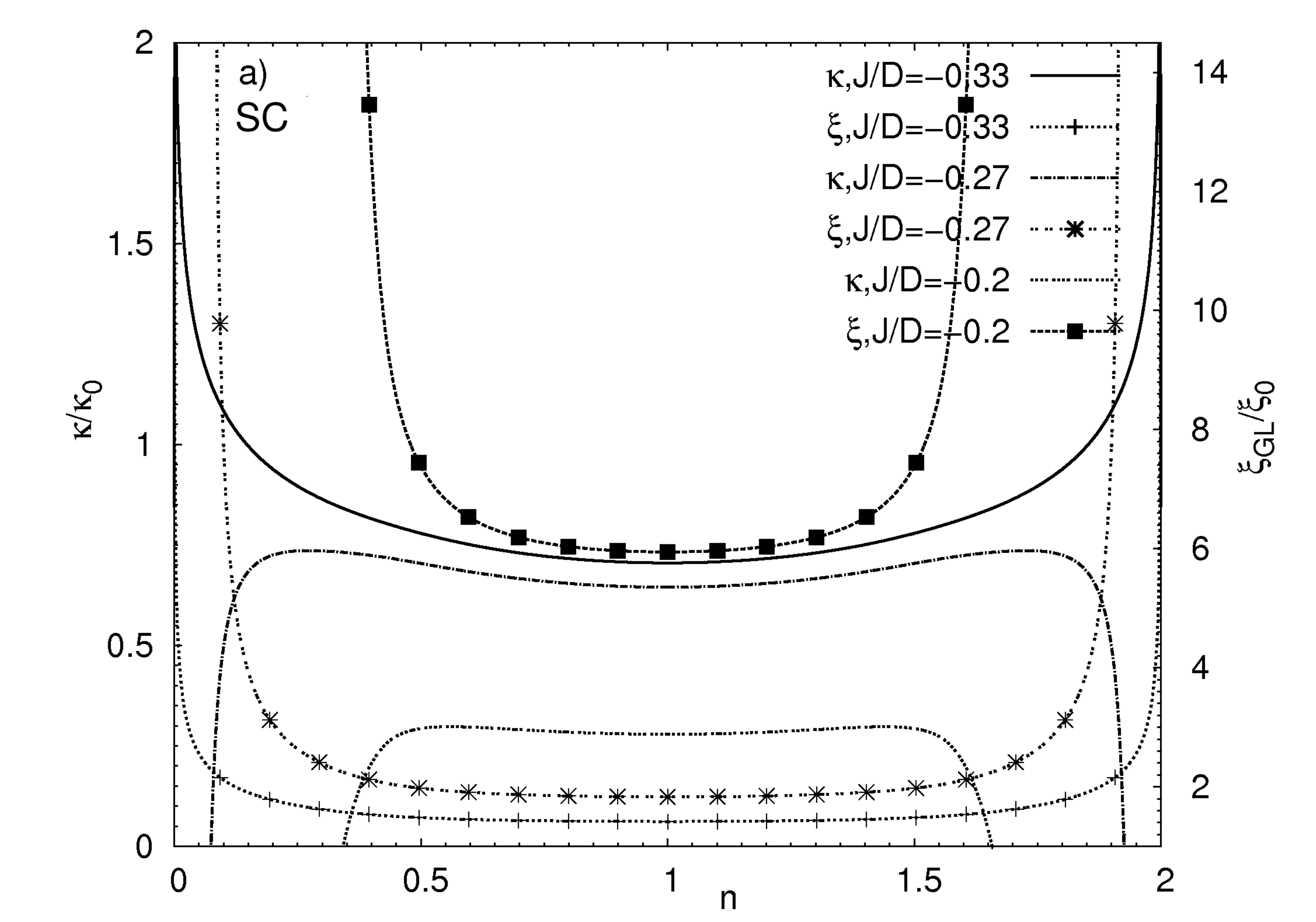}
\includegraphics[width=\sizetwo]{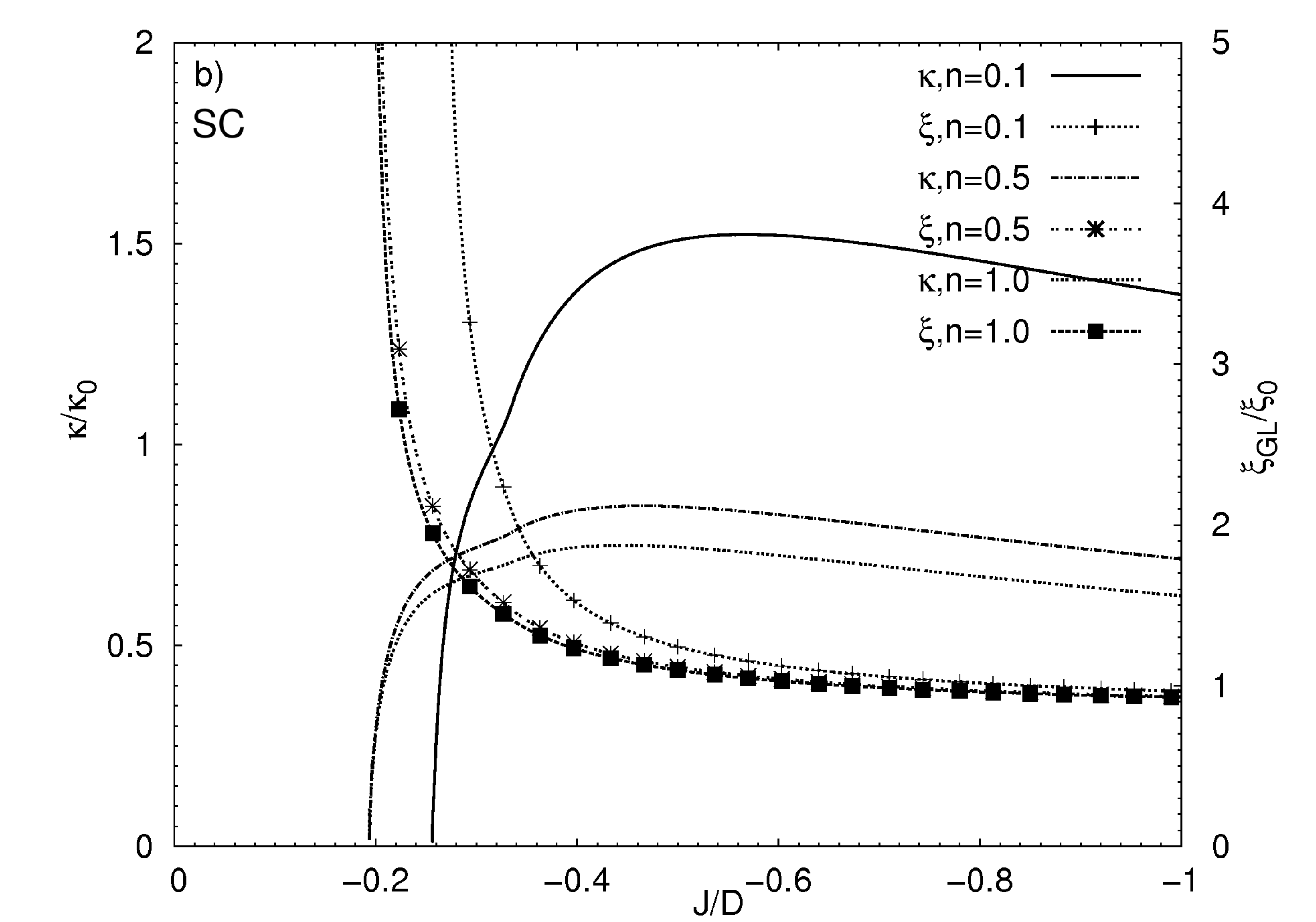}
\caption{
(a) Concentration $n$ and (b) interaction $J$  
dependences of the GL coherence 
length $\xi_{GL}/\xi _0$ ($\xi _0=a/\sqrt{2z}$, lines with symbols) and the 
Ginzburg ratio $\kappa /\kappa_0$, ($\kappa _0=(\hbar c/e)\sqrt{2}/\sqrt{\pi
Da^{4-d}}$, lines without symbols) at $T=0$. 
Panel (a) is plotted for fixed values 
of $J$: $J/D=-0.2$, $J/D=-0.27$, $J/D=-0.33$, while panel (b) 
for fixed $n$: $n=0.1$, $n=0.5$, $n=1$ (for the SC lattice).
}\label{fig:k-xi_vsn-sc}
\end{figure*}

The ground state numerical results for the eta-pairing order parameter $x_\eta$ as  a function of $n$ for several  values of $J/D$ are presented in Figs. \ref{fig:x-Eg-vsn}(a) (SQ lattice) and \ref{fig:x-Eg-vsn-sc}(a) (SC lattice), 
whereas Figs. \ref{fig:x-Eg-vsn}(b) (SQ lattice) and \ref{fig:x-Eg-vsn-sc}(b) (SC lattice) show $x_{\eta}$ as a function of $J$ (for several representative values of $n$). 
Analogous plots for $d=\infty$ lattice (semi-elliptic DOS) are shown in Fig. \ref{fig:x-Eg-vsn-inf}.
The transition from the normal to the eta--phase at $T=0$ is of the second order. 
When the ordered phase sets in for $|J|>|J_c|$, the parameter $x_{\eta}$ continuously increases  till it  attains 
its maximum value (dependent on $n$) in the strong-eta regime. 
In the \emph{strong} eta--pairing regime (i.e., $|J| \geq |J_{c1}|=2t$) the superconducting eta-phase is stable  within the whole range of concentration $n$ ($0<n<2$).
In this regime, as we have mentioned earlier, the magnitude of the order parameter $x_\eta$ assumes its maximum value  $x_\eta ^{\max }=\frac{1}{2}\sqrt{n\left( 2-n\right)}$. 
In the \emph{weak}-eta regime (i.e., $|J| < |J_{c1}|=2t$) the eta--phase occurs only within a limited range of $n$, 
and the range shrinks with decreasing $|J|$. 
The magnitude of parameter $x_{\eta}$ decreases with decreasing $|J|$ and $n$. 
Close to phase boundary the $x_{\eta}$ quickly vanishes with $J\rightarrow J_c$ and with $n\rightarrow n_c$, except for the SQ lattice, where at half-filling, due to van Hove singularity in the DOS, parameter $x_{\eta}$ exponentially vanishes with $J\rightarrow 0$. 
As one can see in the presented plots for SQ and SC lattices, the effects of the the Fock term does not change qualitatively the plots with respect to the case of the model equations without the Fock term. 
The effects vanish in the strong eta-pairing regime as well as at half-filling for SQ lattice when $J\rightarrow 0$. 
Moreover, in the weak-eta phase they increase the value of $x_{\eta}$ and expand the range of $n$ occupied by the eta--phase.



\subsubsection{Superconducting characteristics at $T=0$}

\begin{figure*}[t]
\centering
\includegraphics[width=\sizetwo]{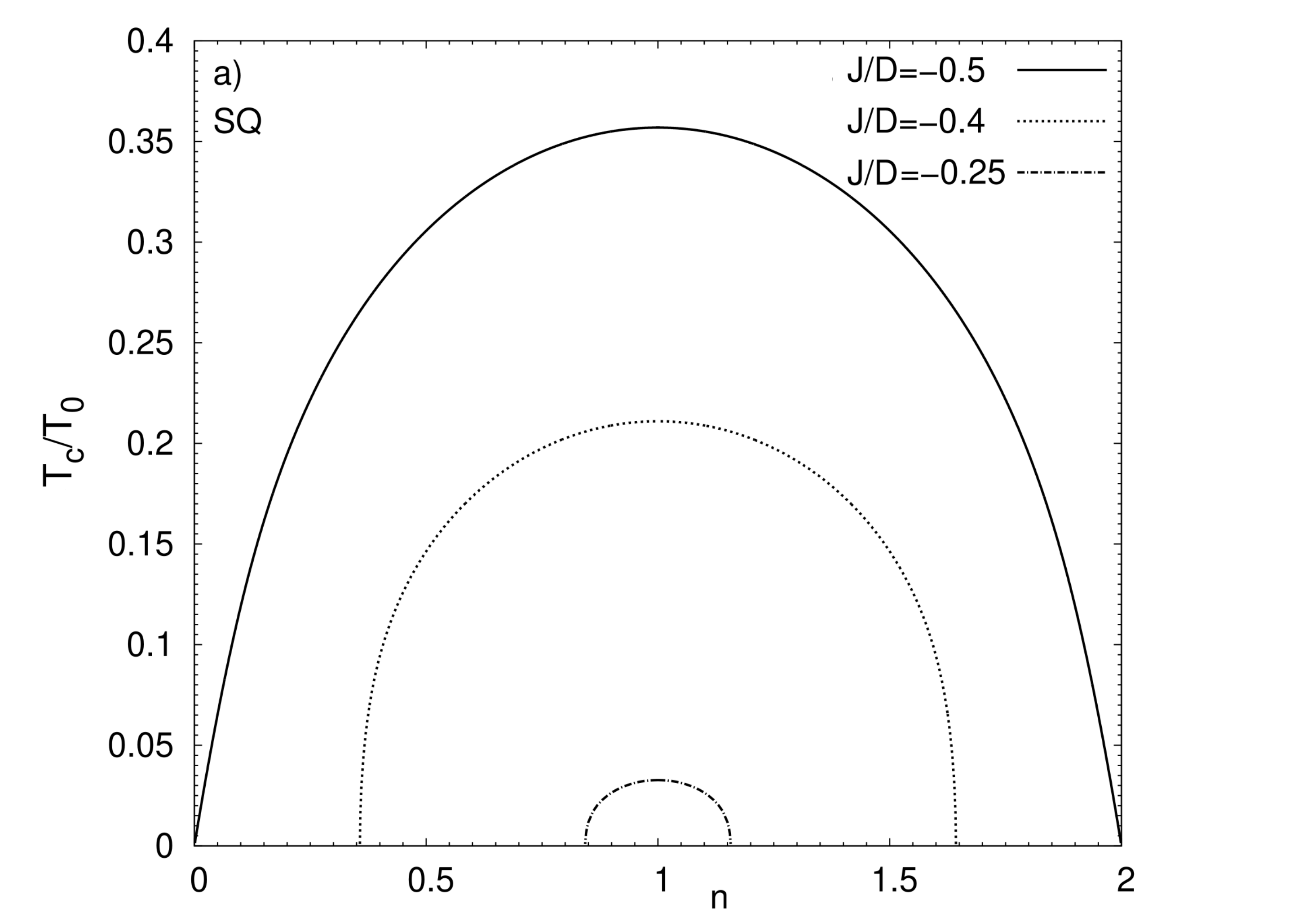}
\includegraphics[width=\sizetwo]{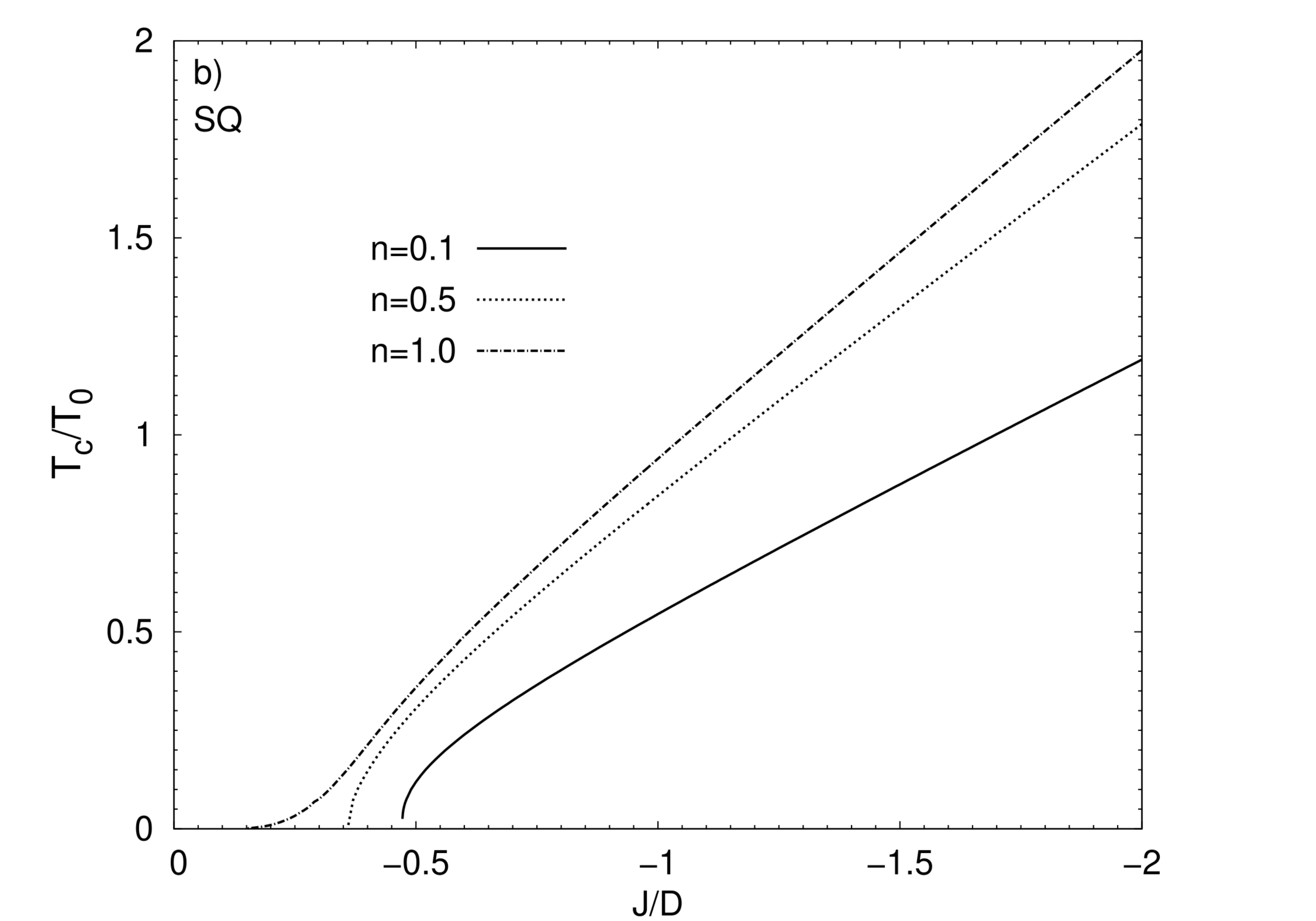}
\caption{
(a) Concentration $n$ and (b) interaction $J$ 
 dependence of the Hartree--Fock critical 
temperature $T_c/T_0$ ($T_0=D/k_B$, results for the SQ lattice, $D=4t$). 
Panel (a) is plotted for fixed values 
of $J$: $J/D=-0.25$; $J/D=-0.4$; $J/D=-0.5$;  
while panel (b) for fixed $n$: $n=0.1$; $n=0.5$; $n=1$. 
}\label{fig:THF-vsn-sq}
\end{figure*}

\begin{figure*}[t]
\centering
\includegraphics[width=\sizetwo]{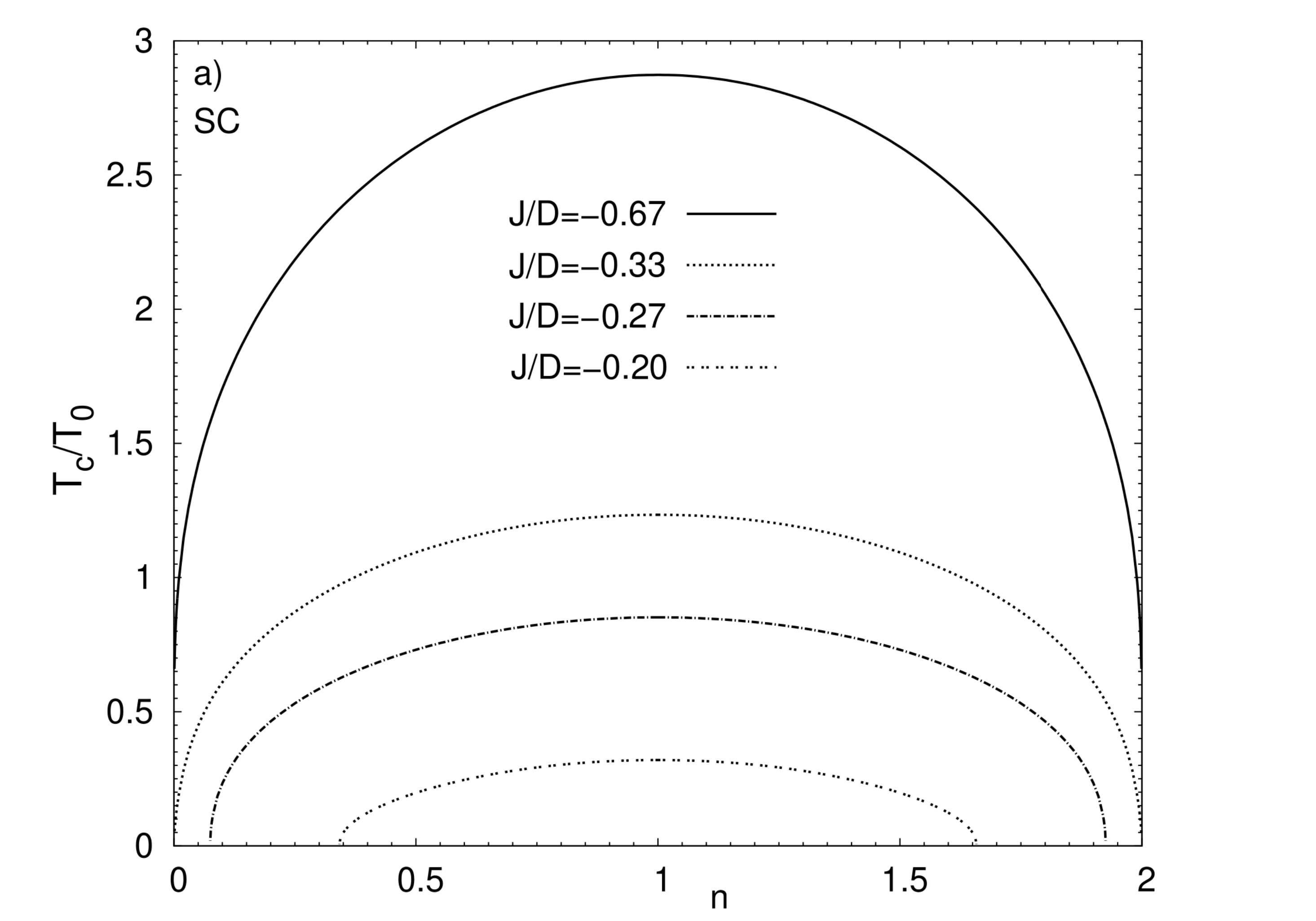}
\includegraphics[width=\sizetwo]{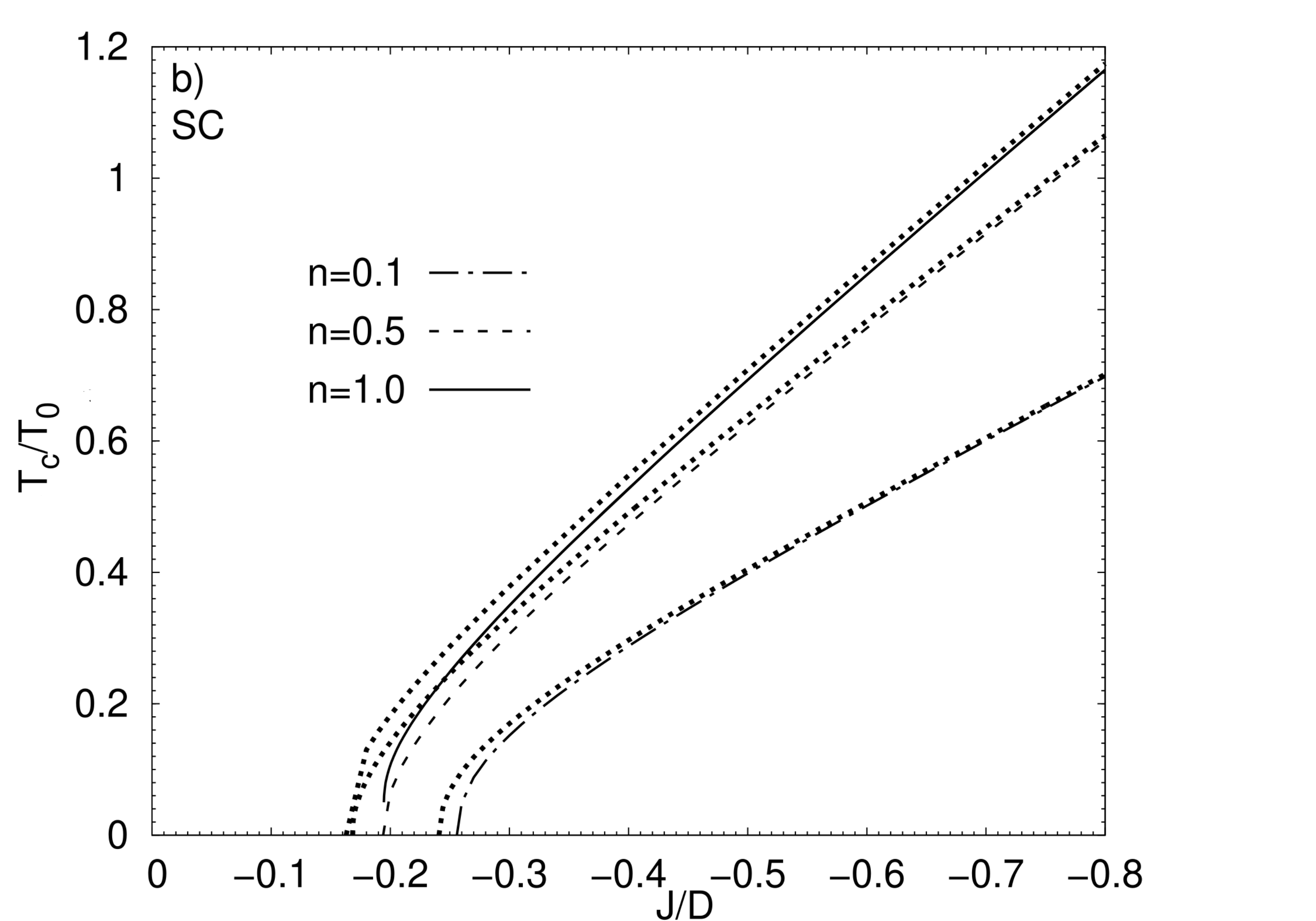}
\caption{(a) Concentration $n$ and (b) interaction $J$ 
 dependence of the Hartree--Fock critical 
temperature $T_c/T_0$ for eta--pairing ($T_0=D/k_B$, results for the SC lattice, $D=6t$).
Panel (a) is plotted for fixed values 
of $J$: $J/D=-0.2$; $J/D=-0.27$; $J/D=-0.33$; 
$J/D=-0.67$, 
while panel (b) for fixed $n$: $n=0.1$; $n=0.5$; $n=1$. 
The lines for the case of the PK model without the Fock term are plotted with a line styles 
defined in the plot legende and respective effects of the Fock term are plotted as dotted lines.}
\label{fig:THF-vsn-sc}
\end{figure*}

In Figs. \ref{fig:l-Hc-vsn}  and \ref{fig:l-Hc-vsn-sc} (for the SQ and SC lattices, respectively) 
the numerical results for the London penetration depth (its inverse square value $1/\lambda ^2$) and the critical field (its square value $H_c^2$) as a function of $n$ and $J$ for several fixed values of $J/D$ ($D=zt$) and $n$, respectively.
%
%
Both $1/\lambda^2$ and $H_c^2$ monotonically decrease with decreasing $|J|$, but the decrease of $1/\lambda^2$ is not smooth as opposed to what has been found for s-wave pairing \cite{czart-11}.
In the case of s-phase both these quantities $1/\lambda^2$, $H_c^2$ evolve smoothly between the limit of weakly interacting single-particle carriers and that of tightly bound pairs for any $n$ \cite{czart-11}.

In the \emph{strong} eta-pairing regime ($|J|>|J_{c1}|$), the $\lambda^{-2}$ and $H_c^2$ are  finite within the whole range of $n$ ($0<n<2$). 
The maximum values of these characteristics are attained at half-filling.
In this  regime $\lambda^{-2}$ and $H_c^2$ behave as $\sim J$ and $\lambda ^{-2}\sim n(2-n)$ for any $n$. 
At low density  limit $\lambda ^{-2}\sim n$. 
Such low density behaviour is similar to that of fermions in the continuum. 
Notice that the strong coupling behaviour of $\lambda ^{-2}$ is similar to that found for s-wave pairing in the PK model with attractive $J$ (see Figs. 1 and 2 of Ref. \cite{czart-11}).

In the \emph{weak} eta--pairing regime, the linear decrease of the $H_c^2$ and $\lambda^{-2}$ 
changes to exponential one ($H_c^2$ and $\lambda^{-2}\sim \exp(|J|/D))$ for the SQ lattice. 
In the case of the SC lattice only $H_c^2$ exhibits the exponential decrease.  
When approaching the phase boundary (between eta and N phase) the $H_c^2$ vanishes,  and $\lambda^{-2}$ takes its lowest but non-zero value.

The evolution of the Ginzburg-Landau (GL) coherence length $\xi _{GL}$ and the Ginzburg ratio $\kappa =\lambda /\xi_{GL}$ with $n$ and $J$ are shown in Figs. \ref{fig:k-xi_vsn} and \ref{fig:k-xi_vsn-sc} (for the SQ and SC lattices, respectively).

Note a substantial variation of $\xi_{GL}$ for both lattices, with $n$ and $J$ when approaching the phase boundary. 
$\xi_{GL}$ takes its minimum value at half-filling. 
The $\xi _{GL}$ decreases exponentially at $J$ close to $J_c$ and goes to a constant value $\xi _\infty =\frac{a}{\sqrt{2z}}$, the same for all $n$, at large $J$. 
For the SQ lattice we observe a visible small irregularity in the monotonous evolution of $\xi _{GL}$  versus $J$ at weak-eta to strong-eta crossover. 
The irregularity is not noticeable in the plot for the SC lattice and does not occur in the case of s-wave  pairing (cf. Figs. 3 and 4 of Ref. \cite{czart-11}).
In the eta-phase, when approaching with $J$ to the phase boundary, the minimum found in the plot $\xi _{GL}$ versus $n$ at $n=1$  sharpens in the case of the SQ lattice, while for the SC lattice for any $J>J_c$ the minimum is more flat.

The evolution of $\kappa$ versus $J$ for SQ lattice in the eta-phase is  presented in Fig. \ref{fig:k-xi_vsn}(b).
With increasing $|J|$ after initial rapid increase in the \emph{weak} coupling regime, away from the low density limit $\kappa$ passes through a round maximum and changes its monotonically at $J=J_{c1}$. 
For low $n$ the Ginzburg ratio exhibits a small irregularity (``bump'') at $J=J_{c1}$ and then passes through a flat maximum. 
For the SC lattice $\kappa$ evolving with $J$ between weak and strong coupling limits undergoes single  
round maximum [cf. Fig. \ref{fig:k-xi_vsn-sc}(b)]. 
Analogously to the s-phase, in the regime $J>>J_{c1}$, $\kappa$ monotonically  decreases as $\kappa \sim \sqrt{1/J}$  for both lattices considered.

The $n$-dependence of $\kappa$ differs substantially for s- and eta- phases.
For s--wave $\kappa$ sharply increases with $|n-1| \rightarrow 1$ for any $J$ \cite{czart-11}, while for 
eta-phase this behaviour is observed for both considered lattices, only in the strong--eta regime. 
In the weak-eta  phase $\kappa (n)$ sharply decreases to zero with $n$ approaching the eta-phase boundaries. 
In this regime we find a round maximum in $\kappa$ versus $J$ plot at $n=1$ for the SQ lattice.
The maximum flattens for $J$ close to $J_{c1}$, and  gradually transforms into a minimum with increasing $J$. 
In the case the of the SC structure $\kappa$ has a minimum at half-filling for any $J$, and in the weak-eta regime flat maximums appear  n the proximity to transition to the N state.
At $T=0$ in the strong coupling limit one finds universal $n$-dependence $\kappa \sim \left[ n\left( 2-n\right) \right] ^{-1/2}$.

\subsection{Finite-temperature diagrams and superfluid characteristics}
\label{sec:resdis-FinTemp}

\begin{figure*}[t]
\centering
\includegraphics[width=\sizetwo]{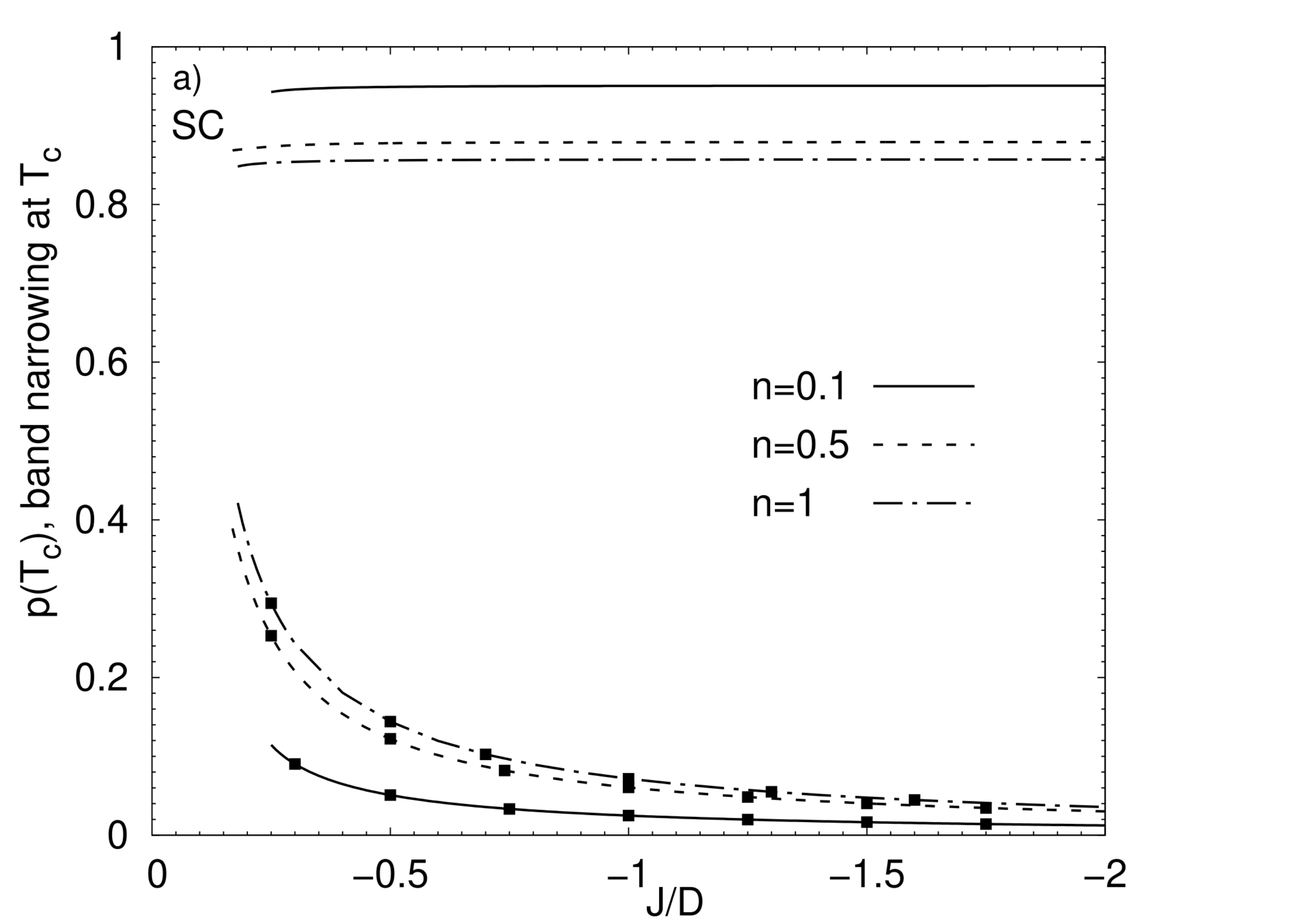}
\includegraphics[width=\sizetwo]{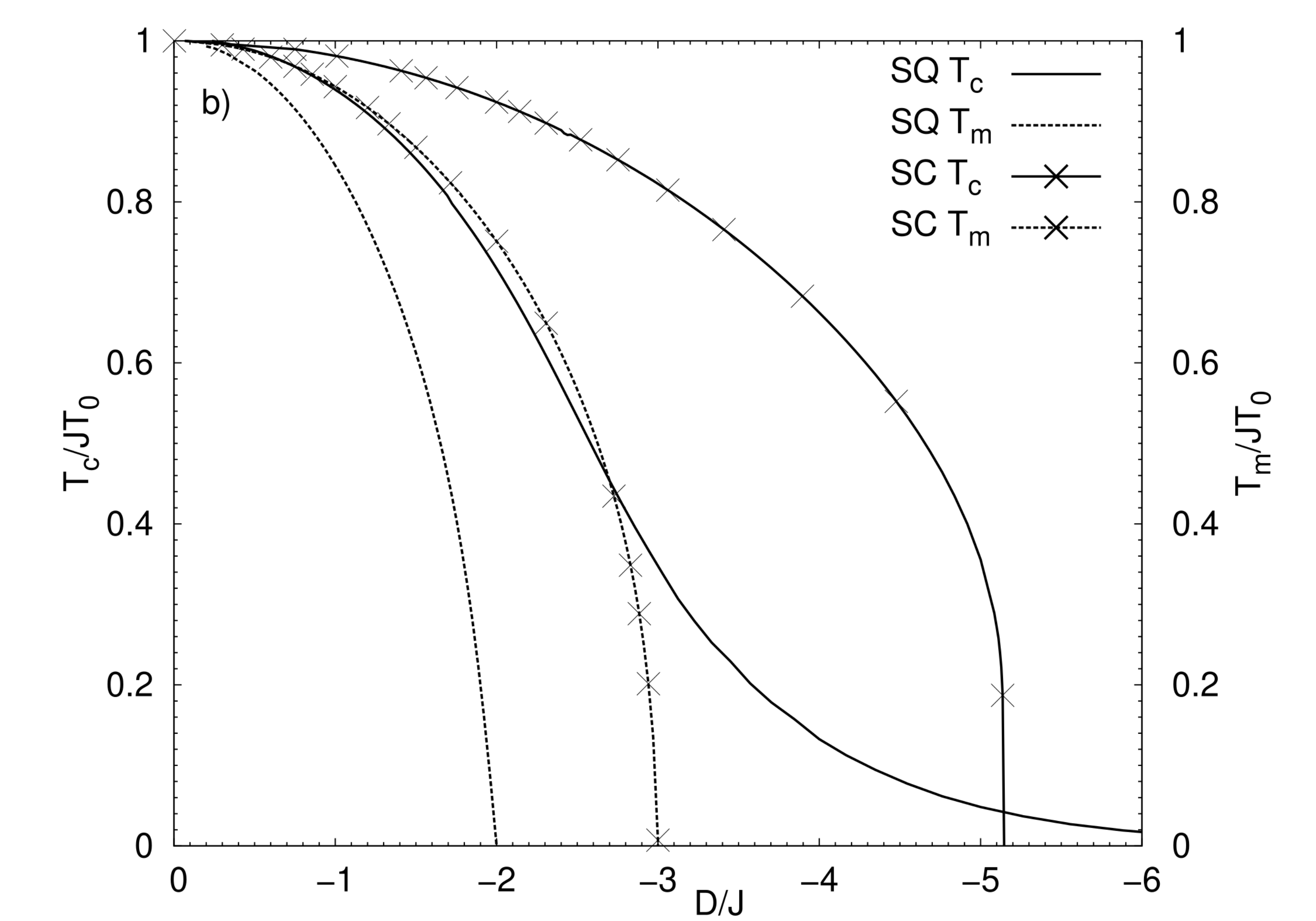}
\caption{(a) The Fock parameter $p(T_c)$ (lower) and band narrowing $1+2p(T_c)J/D$ (upper) at $T_c$ 
for eta--pairing plotted as a function of $J/D$ for $n=0.9999$, $n=0.5$ and $n=0.1$ (plotted for SC lattice, $D=6t$). 
(b) Critical temperatures $T_{c}/T_0$ ($T_0=D/k_B$) for eta--pairing and the temperature $T_m/T_0$ 
 at which $E^{min}_g$ vanishes (lines with symbols), plotted as a function of $D/J$ for $n=1$
(results for the SQ and SC lattices, as labeled).
}
\label{fig:FockandTm}
\end{figure*}

In Figs.  \ref{fig:THF-vsn-sq}, \ref{fig:THF-vsn-sc}, and \ref{fig:THF-vsn-inf} (for the SQ, SC and $d=\infty$ lattices, respectively), we show  the evolution of the Hartree--Fock critical temperature $T_c$ versus $n$ [panels (a)] and versus $J$ [panels (b)]  for a few fixed values of $J$ and $n$, respectively.
In the weak-eta pairing phase the $T_c \neq 0$ is restricted to a limited range of $n$ and the range vanishes with shrinking the eta--phase stability region  with $J\rightarrow J_c$. 
Close to $J_c$ with decreasing $J$ the $T_c$ sharply drops to zero, except for the case of SQ lattice at $n=1$ (in this case the $T_c$ drops to zero exponentially for $|J|\rightarrow 0$)
(cf. Figs. \ref{fig:THF-vsn-sc}b, \ref{fig:THF-vsn-sq}b
and \ref{fig:THF-vsn-inf}b).  
In the strong coupling limit $T_c$ becomes proportional to $J$ for any electron concentration as expected in $t\rightarrow 0$ limit \cite{KRM2012,KR2013}.

Analogously to the behaviour found for $x_{\eta}$ at $T=0$, the effects of the the Fock term on $T_c$ does not change qualitatively the plots of this characteristic with respect to the case of the model equations without the Fock term. 
The strongest influence of the term on $T_c$ is observed for small values of pairing strength, it decreases with increasing $|J|$ and  vanishes for large $|J|$. 
The Fock parameter $p(T_c)$ and band narrowing at $T_c$ for eta-pairing plotted as a function of $J/D$ for several fixed values of $n$ for SC lattice are shown in Fig. \ref{fig:FockandTm}(a). 
%


The energy gap $E_g^{min}$, which exists at $T=0$ for $\left| J\right| >\left| J_{c1}\right|$, is reduced with increasing $T$.
It vanishes at some characteristic temperature which is denoted  here as $T_m$. 
In Fig. \ref{fig:FockandTm}(b), we present the dependencies of $T_c$ and $T_m$ as a function of $D/J$ for the SQ and SC lattices (calculated at $n=1$). 
As we see $T_c>T_m$, except for $D/J=0$ (the atomic limit \cite{KRM2012,KR2013}), and even for the 
\emph{strong} eta-pairing limit, there exists a range of temperatures where the system exhibits a gapless behavior. 
Notice that in the case of the SQ lattice due to van Hove  singularity $T_{c}\rightarrow 0$ at $n=1$ only for $D/J \rightarrow -\infty$.


Decreasing temperature,  the gap in the quasiparticle energy spectrum appears at $T_c$, but  except $d=\infty$ it cannot be associated with any real phase transition to a superconducting state.  
The $T_c$ should be rather treated as a reliable estimate of the pair formation temperature and appearance of the pseudogap \cite{Micnas-90,Singer-98,Singer-98a}.
However, it is by no means a rigorous borderline, since some pairing correlations are present at all temperatures. 
In particular, for $d=2$ SQ lattice  only at the critical temperature of the K--T transition $T_{KT}$, the phase coherence sets in and the transition to a phase with bound vortex--antivortex pairs occurs. 
Thus in the region between $T_c$ and $T_{KT}$ one has a state of incoherent pairs. 
Except for the weak coupling regime $|J/t|<<1$ and $d=\infty $, a strong influence of
the phase fluctuations on the superconducting pairing is found.

In the strong eta-phase regime we estimate the upper bound of the phase transition temperature $T_{KT}$ for $d = 2$ SQ lattice  with the ground state value of superfluid stiffness $\rho_s$ (helicity modulus) Eq. (\ref{TKTtylda}). 
This estimation still allows for  qualitative assessment of the K--T temperature.
With decreasing $J/D$ the phase fluctuations are becoming less significant and the $T_c$ becomes reliable upper bound estimation of the phase transition temperature.

\begin{figure*}[t]
\centering
\includegraphics[width=\sizetwo]{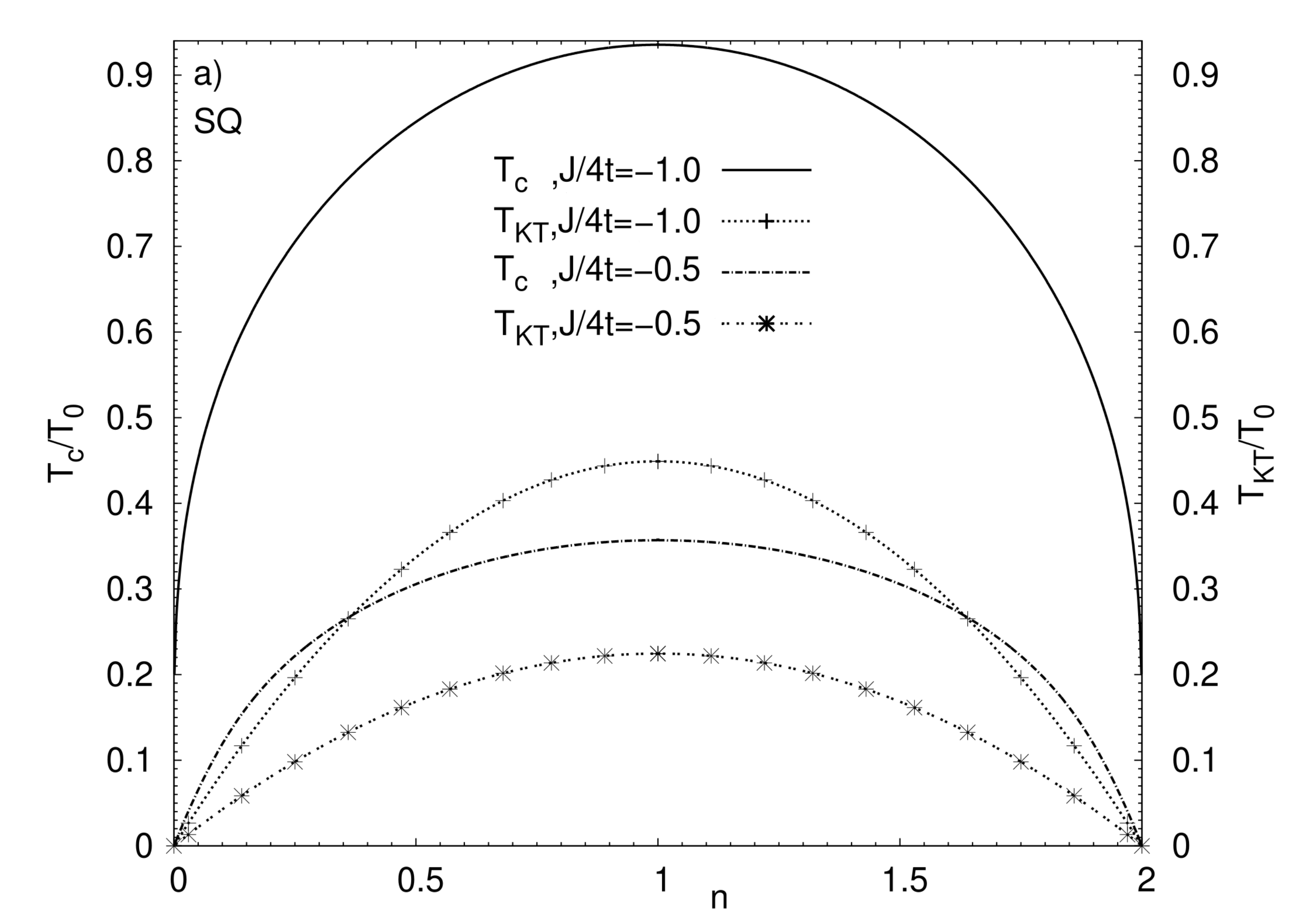}
\includegraphics[width=\sizetwo]{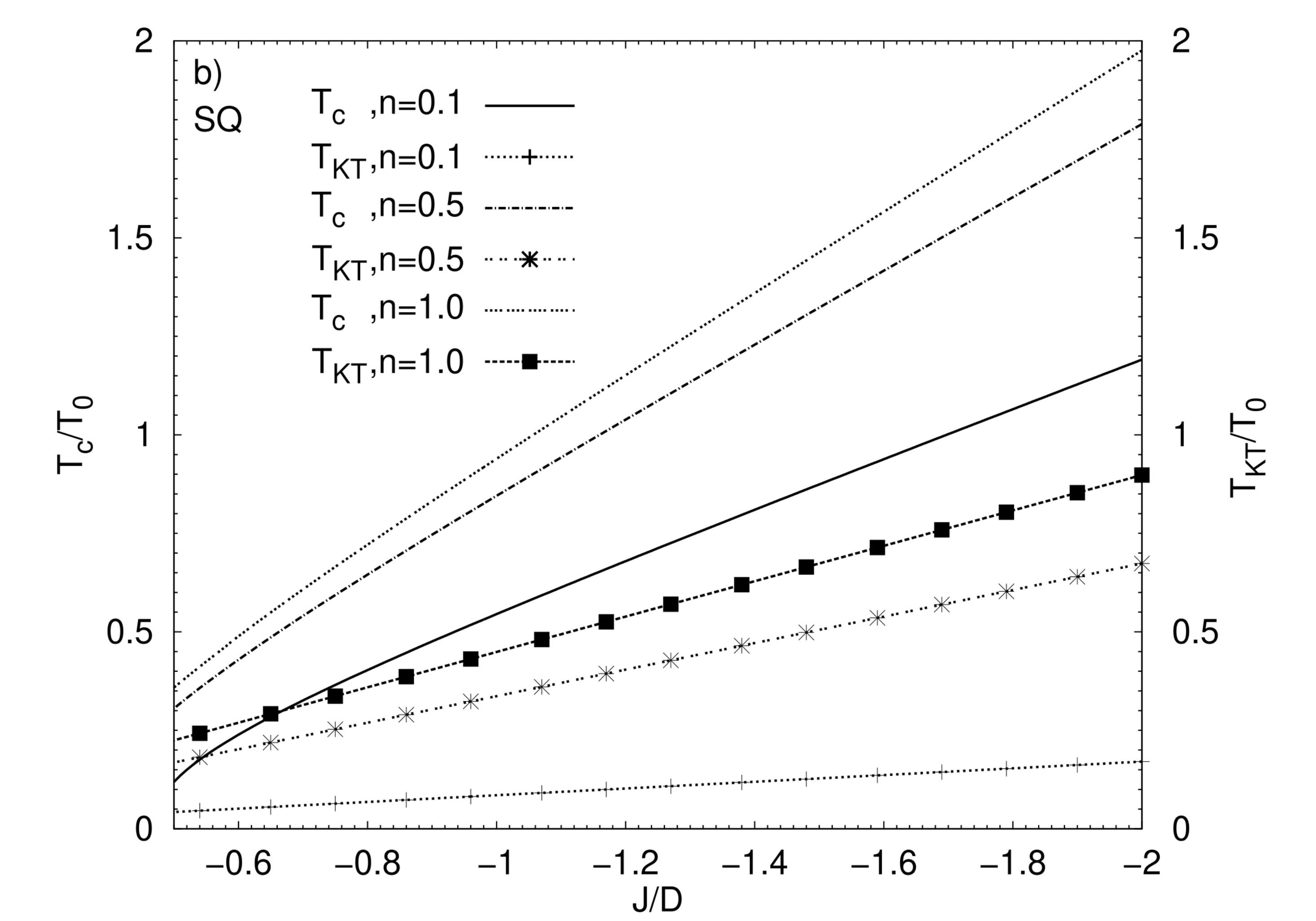}
\caption{
(a) Concentration $n$ and (b) interaction $J$ 
 dependence of the Hartree--Fock critical 
temperature $T_c/T_0$ (lines without the symbols) and upper bound for the Kosterlitz--Thouless critical 
temperature $T_{KT}/T_0$ (lines with the symbols); $T_0=D/k_B$.
Panel (a) is plotted for fixed values 
of $J$:  $J/D=-0.5$; $J/D=-1.0$; while panel (b) for fixed $n$: $n=0.1$; $n=0.5$; $n=1$. 
Results for the SQ lattice.
}\label{fig:THF_TKT_vsn}
\end{figure*}

\begin{figure*}[t]
\centering
\includegraphics[width=\sizetwo]{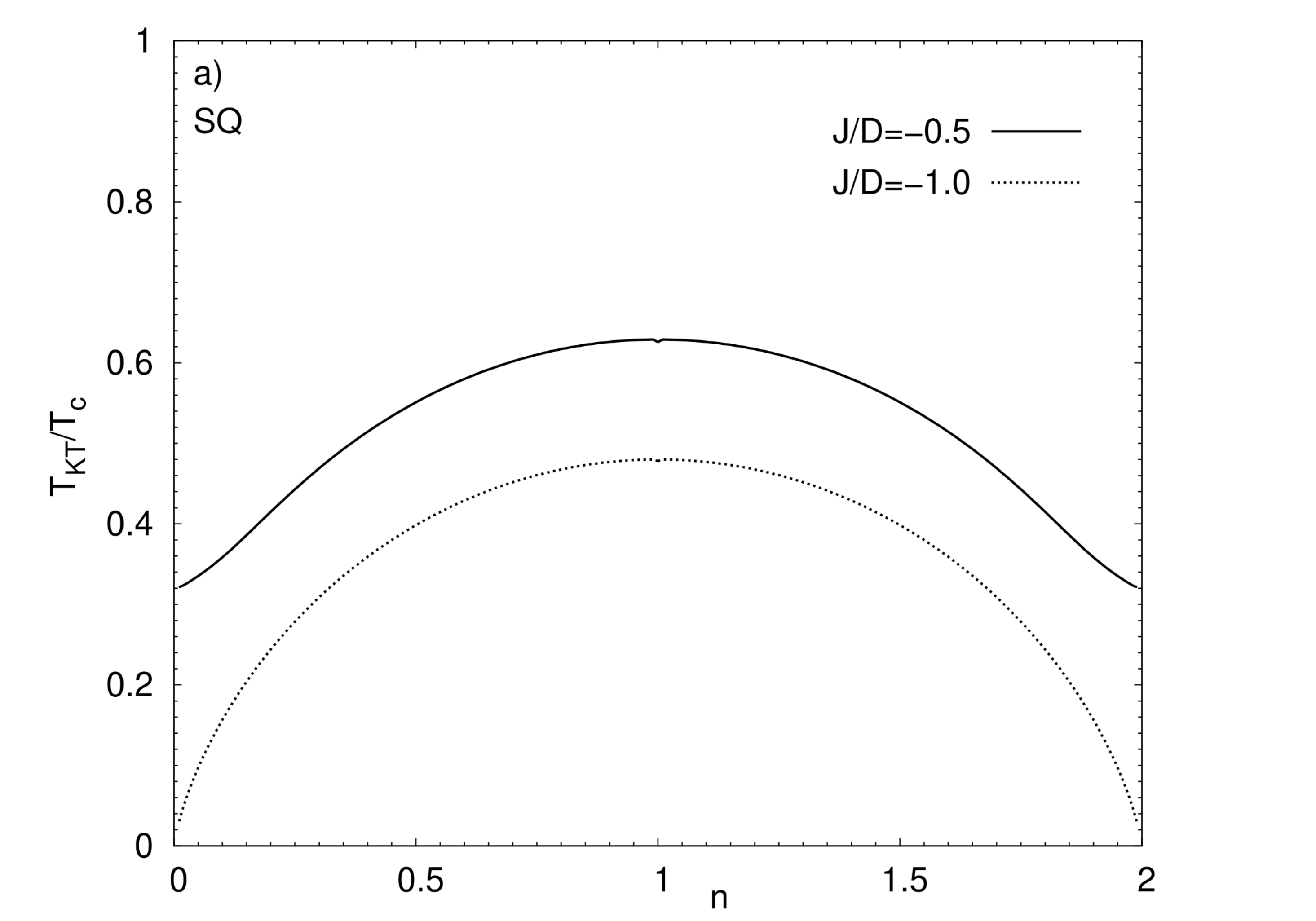}
\includegraphics[width=\sizetwo]{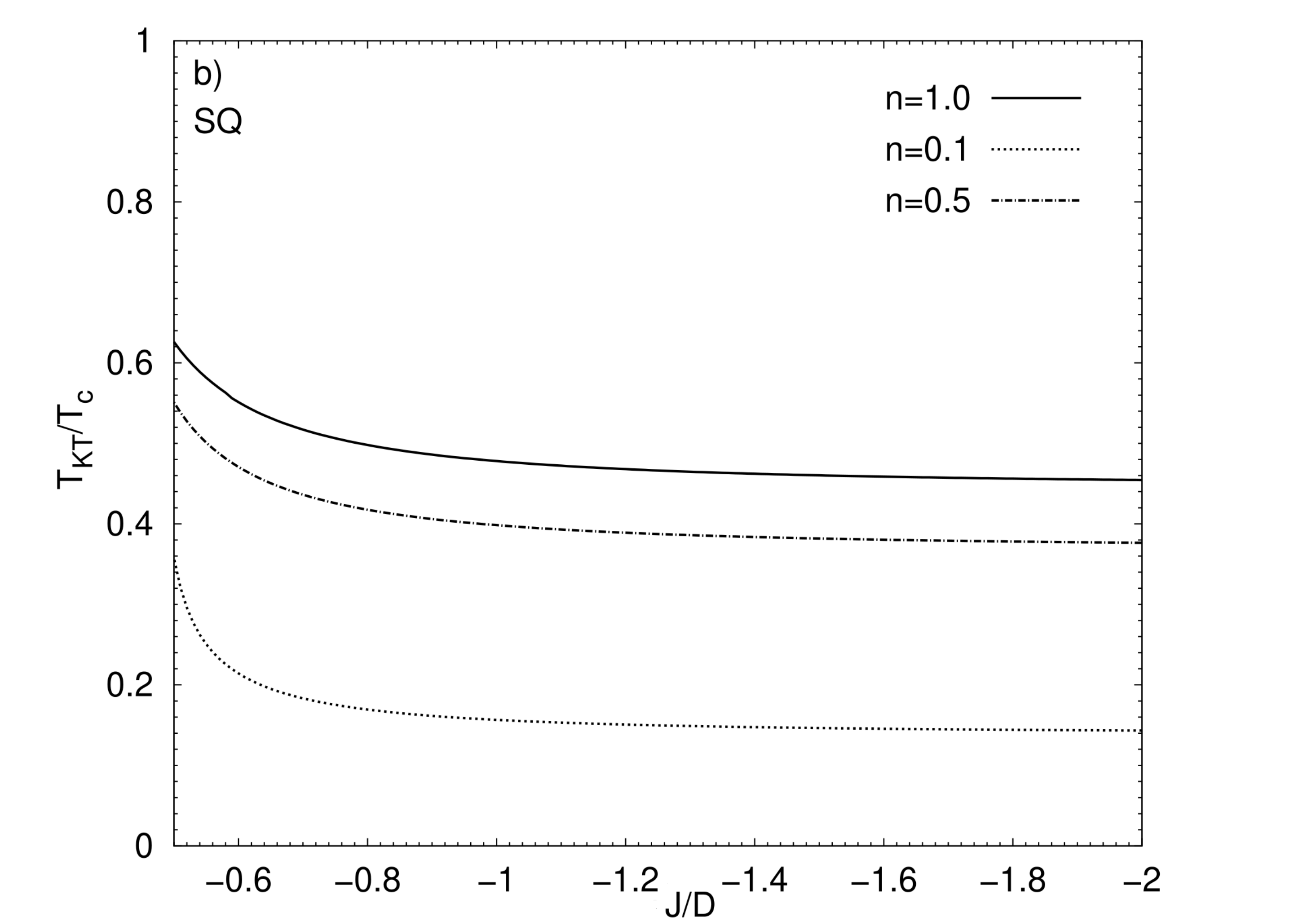}
\caption{
(a) concentration $n$ and (b) interaction $J$  dependencies of the $T_{KT}/T_c$ ratio plotted for 
 fixed values of $J/D=-0.5$; $J/D=-1.0$ [panel (a)] and $n=0.1$; $n=0.5$; $n=1$ [panel (b)]. 
 Results for the SQ lattice.
}\label{fig:TKTByTHF_vsn}
\end{figure*}

In Figs. \ref{fig:THF_TKT_vsn} and \ref{fig:TKTByTHF_vsn} the plots of critical temperatures $T_c$, $T_{KT}$ and $T_{KT}/T_c$ ratios versus $n$ and $J$ are presented for a few fixed $J/D$ and $n$, respectively.
The difference between $T_c$ and $T_{KT}$ attains its minimum at half-filling,  and it increases when moving  away from $n=1$ [Figs. \ref{fig:THF_TKT_vsn}(a) and \ref{fig:TKTByTHF_vsn}(a)]. 
The K--T transition temperature $T_{KT}$ can be much smaller than $T_c$ and the highest 
reduction is found at low carrier concentration.
As we can notice from Figs. \ref{fig:THF_TKT_vsn}(b) and \ref{fig:TKTByTHF_vsn}(b), the effect of the phase fluctuations on the superconducting pairing decreases with decreasing $|J|$. 

In the strong coupling regime the results obtained for the eta-phase are qualitatively similar to those we have found for s--wave pairing (cf. Figs. 7 and 8 of Ref. \cite{czart-11}).
In this limit for both types of superconducting orderings, except for $d=\infty$, there is strong impact on the phase fluctuations on the superconducting pairing.
The $T_{KT}$ is much lower than $T_c$. 
The effects of the phase fluctuations increase with increasing coupling and with decreasing concentration. 
Moreover, for NN hopping only, the $T_c$ and $T_{KT}$ temperatures attain their maxima at $n=1$ and their plots are 
symmetric with respect to the $n\rightarrow 2-n$ transformation.

In the weak coupling limits the behaviour of the $T_{KT}$ and $T_c$ versus the model parameters for both pairing types can be much different. 
In the case of s-phase  both temperatures  are finite for any $n$ ($0<n<2$), while in the case of the eta-phase $T_{KT}$ and $T_c$ versus $n$ plots are restricted to a limited range of $n$. 
Moreover, in the eta-phase the both temperatures vanish with decreasing $J$ at finite value $J_c$, while for s-phase they are finite for any $J$ ($0<|J|$).

\section{Concluding remarks}
\label{sec:conclusions}

The PK model can be considered as a simple effective model for description of  superconductors with short coherence length  and with s--wave ($J>0$, condensate of electron pairs with $\vec{q}=0$) and eta--pairing ($J<0$, condensate of electron pairs with $\vec{q}=\vec{Q}$), including those which form isotropic $d=3$ structures as well as those with layered (quasi $d=2$) structures. 
For the model the HFA is a rigorous theory in the $d \rightarrow \infty$ limit.

For $J<0$, the superconductivity is due to the eta-pairing mechanism, where the on-site singlet pairs display off-diagonal long-range correlation with phase $\pi$. 
However, in the studied model, the pairing interaction is in the form of intersite pair-hopping $J$ interaction, not of the on-site or inter-site density-density attraction as in (extended) Hubbard models \cite{Micnas-90,Robaszkiewicz-99}. 
The eta-pairing can be also treated as a very peculiar case of the Fulde-Ferrell-Larkin-Ovchinnikov phase with the largest possible $\vec{q}=\vec{Q}$ at the vertex of the first Brillouin zone \cite{PtokPRA2013,PtokPRA2017}. 
Note also that the eta-pairing states were originally introduced for the mathematical purpose to solve the Hubbard model analytically \cite{YangPRL1989}.

The properties of the PK model with repulsive $J$ which favours eta-pairing are qualitatively different from those of the model with attractive $J$ which stabilizes s--wave. 
In the case of s-wave pairing (for $J>0$) superconducting characteristics evolve smoothly between the limit of weakly interacting single--particle carriers and that of tightly bound pairs \cite{czart-11}. 
The system in the eta-state  never exhibits standard BCS--like behaviour found in limit of the weakly interacting single-particle s-wave carriers.
As we have found, in contrast to the case of the isotropic s--wave state, the transition into the eta--phase  (for $J<0$) occurs only above some critical value $|J_c|$  (except for the SQ lattice at $n=1$). 
The critical value $J_c$ depends on the lattice structure, i.e., on the form of the density 
of states DOS, $D(\epsilon )$, as well as on the band filling $n$.
Thus, for a given lattice structure  in a definite range of interaction parameter $J/t$ the 
transition from the superconducting eta--phase to the N state can be realized by changing $n$.

The second characteristic value $J_{c1}/t$,  at which $E_g^{min}$ goes to zero, marks the transition between 
the \emph{strong} and \emph{weak} eta-pairing phases.
At $T=0$, the \emph{strong} eta-phase is characterized by the gap between the lower and higher quasiparticle
band $E_g^{min} > 0$ and the order parameter taking its maximal value $x_\eta^{\max}$, whereas in the  \emph{weak} eta-pairing phase $E_g^{\min }<0$ and $0< x_\eta <x_\eta ^{\max }$. 
For $d$-dimensional hypercubic lattices  with NN hopping only, $\left| J_{c1}\right| =2t$ for any $n$.

As we have found in the presented phase diagrams in a certain range of interaction parameter $J$ the eta-phase to the normal phase transition can be realized by changing $n$, and the strongest $n$-dependence is observed for the SQ lattice.
Due to the van Hove singularity at $n=1$ in the SQ lattice $J_c=0$ and  the eta-pairing phase can be stable for any $J<0$. 
In contrast, for the SC lattice and $d=\infty$ structures, $|J_c|$ significantly depends on concentration only for low $n$ and never $J_c\rightarrow 0$. 
For $d$-dimensional hypercubic lattices with NN hopping only (except for Gaussian DOS for the $d=\infty$ lattice), 
the second characteristic value of $J < 0$, $\left| J_{c1}\right| =2t$ for any $n$.
On the presented diagrams we have also found  the location of the crossover to BEC  regime and have shown that the  crossover can be achieved by changing (decreasing) the electron density ($n$ or $2-n$).

In the weak coupling regime the differences in thermodynamic and electrodynamic properties between s- and eta- pairing,  are particularly pronounced.
The weak eta-phase  exists  only  within a restricted range of $n$,  while in the AH with $U<0$ and in the PK model with $J > 0$,   the s--wave pairing superconductivity can be stable at $T=0$ within the entire range of $n$ for any coupling strength.

In the eta-phase the ground state inverse square value of the London penetration depth  ($\lambda^{-2}(T=0)$) and $H_c^2$ monotonically decrease with decreasing $|J|$. 
In the \emph{weak} eta-pairing regime, for the SQ lattice the linear decrease of the both characteristics  changes to exponential one ($H_c^2$ and $\lambda^{-2}\sim \exp(|J|/D))$.
While in the case of the SC lattice only $H_c^2$ exhibits exponential decrease.
The strong coupling behaviour of $\lambda ^{-2}$ and $H_c^2$ are similar to that found for s-wave pairing in the PK model with attractive $J$ (see also Figs. 1 and 2 of Ref. \cite{czart-11}).
In this regime these characteristics are finite within the whole range of $n$  ($\lambda ^{-2}\sim n(2-n)$ for any $n$) and proportional to $J$. 
They achieve the maxima at half-filling and at low density are proportional to $n$ (similar behaviour to that found for fermions in the continuum).

As concerns the coherence length  $\xi_{GL}$ (cf. Fig 3 and 4 of Ref. \cite{czart-11}),  it substantially increases when  approaching the phase boundary, attains its minimum at half-filling [see Figs \ref{fig:k-xi_vsn}(a) 
and Fig. \ref{fig:k-xi_vsn-sc}(a)], and  tends to a constant value $\xi _\infty =\frac{a}{\sqrt{2z}}$, the same 
for all $n$, at large $J$. 
For the SQ lattice in the eta-phase we observe a visible small irregularity   in the monotonous evolution of $\xi _{GL}$  versus $J$ at weak-eta to strong-eta crossover. 
In the eta-phase for the SC lattice the Ginzburg ratio, evolving with $J$ undergoes  a single round maximum, similarly as for $J>0$ \cite{czart-11}.  
For the SQ lattice the evolution of $\kappa$ with $J$ is qualitatively different  from the evolution for the
SC lattice and for the s-wave phase (cf. Figs. \ref{fig:k-xi_vsn}(b), \ref{fig:k-xi_vsn-sc}(b) and Fig. 4 of Ref. \cite{czart-11}).  
In the case of the SQ lattice a minimum on $\kappa$ versus $J$ plot appears at the weak to strong eta-phase crossover.

In our studies we have demonstrated that in the \emph{strong} coupling limit $\left| J\right| \gg 2t$
the electromagnetic and thermodynamic properties of the eta-phase (for $J<0)$ become similar to those of the s-phase (for $J>0$).
In this regime the $H_c^2,$ $1/\lambda ^2\left( 0\right) $ and $T_c$, as well as $T_{KT}$,  become proportional to $\left|J\right|$,  the coherence length $\xi _{GL}$ tends to a constant value $a/\sqrt{2z}$,  while the Ginzburg ratio 
 $\kappa =\lambda /\xi _{GL}\sim 1/\sqrt{\left| J\right| }$,  and the energy gap $E_g^{\min }\Rightarrow \left|
zJ\right|$, for any $n$. 
With decreasing $\left| J\right| /D$, $\xi_{GL}$ increases and becomes $n$-dependent, going to
infinity at $J_c$, i.e., at the border with the N state. 
One finds also a few universal $n$-dependences at $T=0$: $H_c^2\sim n\left( 2-n\right) $ and $\kappa \sim
\left[ n\left( 2-n\right) \right] ^{-1/2}$, for $\left| J\right|\gg 2t$, and $\lambda ^{-2}\left( 0\right) \sim n\left( 2-n\right)$, for any $\left|J\right|>2t$.

We found, that for the PK model the effects of the the Fock term did not change qualitative characteristics of the analysed ground state phase diagrams and the parameters' evolutions plots with respect to the case of the model equations without the Fock term. 
Taking into account of the Fock term expanded the eta-phase stability towards the lower values of $|J|$, and 
in the concentration space towards limits of occupancy. 
In the weak-eta regime, it increased the value of the $x_\eta$ as well as
the value of $T_c$ especially for small values of pairing strength. 
The influence of the Fock term on $x_\eta$ disappeared at strong eta-phase, and for $T_c$ rapidly decreased with increasing $|J|$.

By taking into account the phase fluctuations, one finds that in the $d=2 $ PK model, for both considered pairing types (eta and s), a new disordered phase between $T_c$ and $T_{KT}$ is possible,  analogous to that found in the AH model \cite{Micnas-90,Denteneer-93,Denteneer-93a,Singer-98,Singer-98a} and in the models with intersite density--density attraction \cite{Chattopadhyay-97,Chattopadhyay-97a,Micnas-99,Tobijaszewska-99}. 
In this state  a gap opens up in the fermionic spectrum, but pairs are phase disordered.
Whereas, the phase transition to the superconducting states takes place at $T_{KT},$ at
which the phase coherence sets in.
As we found, in the strong-eta coupling regime, the influence of phase fluctuations increase with decreasing 
concentration. 
Moreover, for both types of pairing in the PK model, the importance of the phase fluctuations increases with increasing coupling (for s--wave cf. Ref. \cite{czart-11}).

Considerations of the effects of on-site $U$ on the superfluid characteristics of the model considered show 
that attractive $U$ ($U<0$) expands the range of stability of eta-phase at $T=0$ towards lower values of $|J|$ \cite{czart-15}. 
Both the s-phase and the eta-phase can survive also for repulsive values of $U$ ($0<U<U_c$).

In Figs. \ref{fig:THF_TKT_vsn} and \ref{fig:TKTByTHF_vsn}, for NN only hopping ($t_2=0$), $T_{KT}$ and $T_c$ are maximal at $n=1$ and their plots are symmetric with respect to the transformation $n\rightarrow 2-n$.
The next-nearest neigbour hopping $t_2\neq 0$ breaks this symmetry and shifts the maxima of these critical temperatures  towards $n<1$ $(n>1)$ for $t_2<0$ $(t_2>0)$ (cf. Fig. 3 for eta-phase and Fig. 2 for s--wave pairing in Ref. \cite{czart-19}).
Moreover, $t_2$ can yield a substantial enhancement of the maximal values of $T_{KT}$ and $T_c$ with respect to the case $t_2=0$, both for $s$-wave pairing  as well as for eta-pairing \cite{czart-19}.

For $t_{ij}=0$ increasing repulsive $U$ changes the nature of the superconducting transition type from a continuous to a discontinuous one, resulting in the tricritical point. 
It also suppresses superconductivity for low $|n-1|$, and causes the system to remain in a normal state for $U/|zJ|>1$ at any $T$ and $n$ (cf. Fig. 2 in Ref. \cite{czart-15} as well as Refs. \cite{KRM2012,KR2013}).

In order to further extend presented in this paper results  in our forthcoming work we are going to present 
investigation of the eta-superconducting ordering and its competition with the magnetic phases in the Penson-Kolb-Hubbard model for  repulsive $J$ ($J<0$) and arbitrary electron density \cite{Robaszkiewicz-99,Japaridze-97,
Mierzejewski-04,Kapcia-15,Kapcia-15a}. 
As it has been already done for the case of s-wave superconducting phase \cite{czart-12,CzartJSNM2019}, in the further work we will analyse the effects of the on-site Coulomb interactions $U$ on the ordered phases stability and eta-phase characteristics on the 2D, 3D and infinite dimensional lattices.
Some preliminary results in the subject have been presented in Refs. \cite{czart-15,KapciaAPPA2018}.

\appendix{}
\section{Explicit expressions for weak and strong couplings}
\label{sec:appA}

From Eqs. (\ref{eqVeta}) and (\ref{eqneta}) one gets
\begin{equation}
\mu = \frac{1}{2} z|J| (n-1), \quad \textrm{for any} \quad  x_\eta \neq 0
\end{equation}
and, in such a case, Eqs. (\ref{eqneta}) and (\ref{eqpeta}) reduce to: 
\begin{eqnarray}
\frac{4X}{z|J|}= \frac{1}{N} \sum_{\vec{k}} \left[\tanh 
\left( \frac{\beta }{2} D_{\vec{k}}^+ \right) - \tanh \left( 
\frac{\beta }{2} D_{\vec{k}}^- \right) \right], \qquad  \label{Xred} \\
p = \frac{-1}{8N} \sum_{\vec{k}} \gamma_{\vec{k}} \left[\tanh 
\left( \frac{\beta  }{2} D_{\vec{k}}^+ \right) + \tanh 
\left( \frac{\beta }{2} D_{\vec{k}}^- \right) \right], \qquad   \label{pred}
\end{eqnarray}
where $D_{\vec{k}}^{\pm} = \epsilon_{\vec{k}} \pm X$ and 
\begin{equation} \label{eq:Xeap}
X = \frac{1}{2} z|J| \sqrt{(1-n)^2+4x_\eta^2}.
\end{equation}
The minimum gap in the spectrum is
\begin{equation}
E_g^{min}=2X-B ,
\end{equation}
where $B=2zt$ denotes the effective bandwidth.

The equation determining $T_c$ ($x_{\eta}\rightarrow 0$ limit) has the form:
\begin{equation}
|n-1| = \frac{1}{2N} \sum_{\vec{k},r=\pm 1} r \tanh  \left[ \frac{\beta_{c}}{2}
\left(\epsilon_{\vec{k}} + \frac{r}{2} z|J| |1-n|\right) \right], \qquad  \label{n_123}
\end{equation}
where $\beta_c = (k_{B}T_{c})^{-1} $ with
\begin{equation}
p=\frac{-1}{8N}\sum_{\vec{k},r=\pm 1} \gamma_{\vec{k}} \tanh\left[ \frac{\beta_c}{2}\left(\epsilon_{\vec{k}} + \frac{r}{2} z|J| |1-n|\right) \right] ,
\end{equation}
while the temperature $T_m=(k_{B}\beta_{m})^{-1}$ at which $E_g^{min}\rightarrow 0$  is determined by
\begin{equation}
\frac {2B}{z|J|}= \frac{1}{N} \sum_{\vec{k},r=\pm 1} r \tanh \left[ \frac {\beta_m}{2}
\left( \epsilon_{\vec{k}} + \frac {rB}{2} \right)\right] \;,
\end{equation}
for $T_m \leq T_c$. 

The ground state  energy of the eta--phase (i.e., at $T=0$) is derived as
\begin{eqnarray}
E_0^{\eta}& = & F_{\eta}\left( \beta \rightarrow \infty \right) \nonumber \\
& = & \frac{1}{2} z|J| (n-1)^{2} + \frac{4}{z} Jp^{2}  + z|J| x_{\eta}^{2} \nonumber \\
& - & \frac{1}{2N} \sum_{k} \left( \left| \epsilon_{k}+X \right| + \left| \epsilon_{k} -
X \right|\right),
\end{eqnarray}
where  $x_\eta$, $p$, and $X$, are given by Eqs. (\ref{Xred})--(\ref{eq:Xeap}) taken for $\beta
\rightarrow \infty$.
At $T=0$ for $|J|>2t$ (the strong eta-phase) one obtains:
\begin{eqnarray}
&& X=\frac{1}{2} z|J|, \ x_{\eta}=\frac{1}{2} \sqrt{n(2-n)}, \ p=0,  \label{X_} \\
&& E_g^{\mbox{min}}=z|J|-B, \nonumber \\
&& E_0^{\eta} = -\frac{1}{4} z|J| n(2-n), \label{X_Eg}
\end{eqnarray}
whereas for $|J|<2t$ (the weak eta-phase) one gets: 
\begin{eqnarray}
E_g^{\mbox{min}}&<&0 , \nonumber \\ 
E_0^{\eta}&=&-\frac{X^2}{z|J|}-2 \int_X^{B/2}
\epsilon D(\epsilon) d\epsilon  \\
&  + & \frac{1}{4} z|J|(n-1)^2 
+ \frac{4}{z} Jp^{2},  \nonumber
\end{eqnarray}
where
\begin{eqnarray}
\frac{2X}{z|J|}&=&\int_{-X}^{X} d\epsilon D(\epsilon), \\ 
 x_{\eta}&=&\sqrt{\frac{X^{2}}{z|J|^{2}}-\frac{1}{4} (n-1)^2 }.
\label{x-weak}
\end{eqnarray}

For the strong eta-phase (i.e., $|J|>2t$) one gets $K_t^{dia}(T=0)=0$, whereas for weak eta-phase (i.e., for $|J|<2t$) one derives
\begin{eqnarray} 
K_t^{dia}(T=0) &= & \frac{8\pi e^2}{\hbar^2c^2a}\frac{ \langle E_{kin} \rangle }{z},\\
\langle E_{kin} \rangle &= &-2 \int_{X}^{B/2} \epsilon D(\epsilon) d\epsilon ,
\end{eqnarray}
and $X$ is given by Eq. (\ref{X_}).
Moreover, one finds
\begin{equation}
K_J^{dia}(T=0)=-\frac{32\pi e^2}{\hbar ^2c^2a} \frac{|J_0|}{z} x_{\eta}^{2} ,
\end{equation}
where $x_\eta $ is given by Eq. (\ref{X_}) for $J>2t$ and by Eq. (\ref{x-weak}) for $J<2t$.

It is worth to note that above expressions agree very well with the corresponding results derived within the MFA (and
also the RPA) for the model of hard--core charged bosons on a lattice (in the
case of absence of intersite boson repulsion) \cite{Micnas-95}.

In the limit of tightly bound pairs (i.e., $t=0$), Eqs. (\ref{eqVeta})--(\ref{eqpeta}),
 (\ref{lambda})--(\ref{Hcxi}), and (\ref{gs}) can be solved analytically at $T=0$ for arbitrary
electron concentration (and any $d$), and the results are: 
\begin{eqnarray}
& & x_{\eta} = \frac{1}{2}\sqrt{n(2-n)}, \quad \mu_{\eta}=\frac{1}{2}J_{0}(1-n), \quad p=0, \nonumber \\
& & \lambda ^{-2} =\frac{8\pi e^2}{\hbar^{2}c^{2}a}Jn(2-n), \frac{H_{c}^{2} a^{d}}{8\pi }  =  \frac{1}{4}zJn(2-n), \qquad \label{eq:propTBpairs} \\
& & \xi _{GL} =\frac{a}{\sqrt{2z}},\quad \kappa =\frac{\hbar c}e\left[ \frac{J}{z} \pi an(2-n) \right]^{-\frac{1}{2}}. \nonumber 
\end{eqnarray}

\begin{figure*}[t]
\centering
\includegraphics[width=\sizetwo]{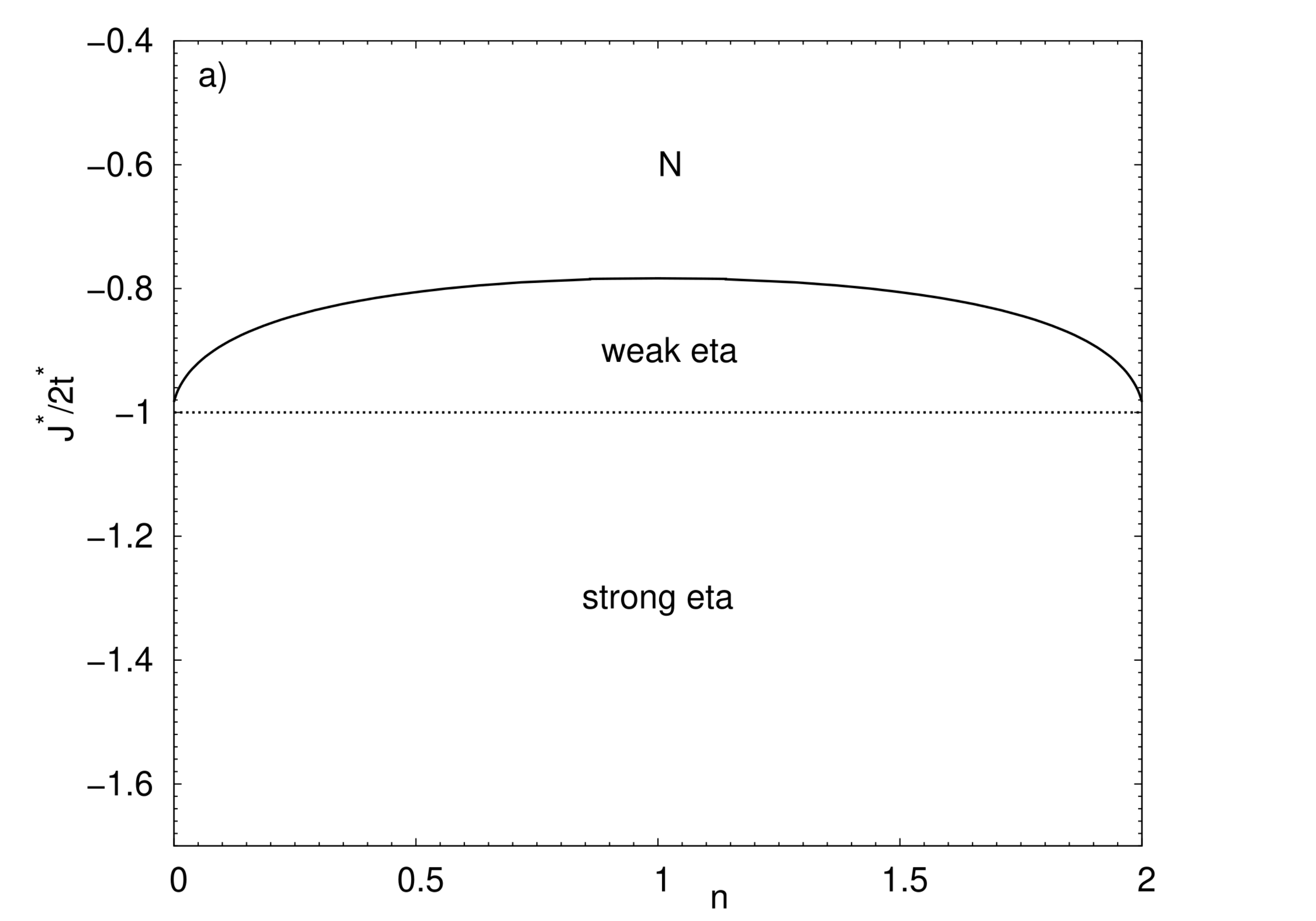}
\includegraphics[width=\sizetwo]{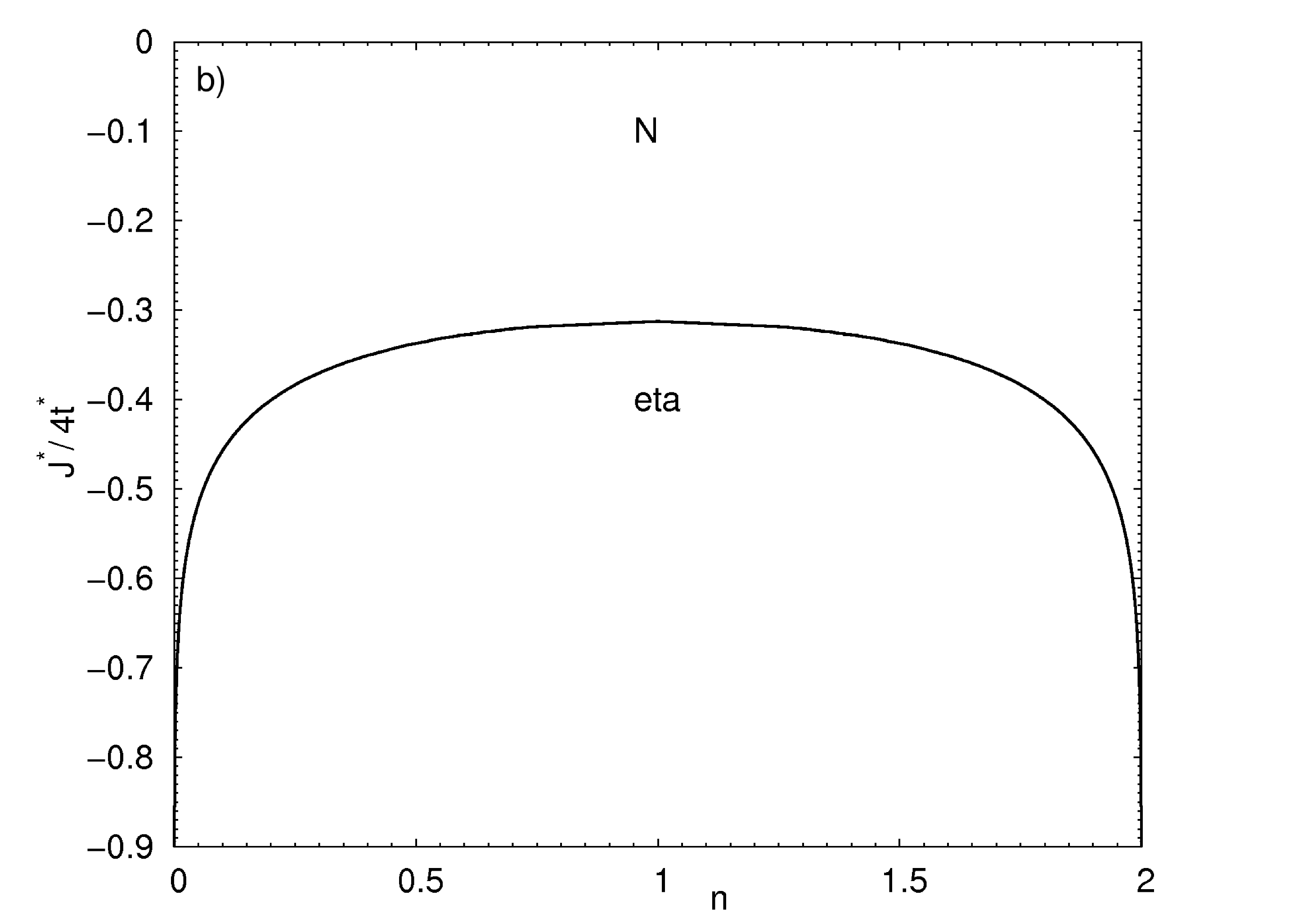}
\caption{
The ground state phase boundaries between the eta--phase and the normal (N) state plotted for $d=\infty$ lattices.  
The semi-elliptic (a) and the  Gaussian (b) density of states used.
}\label{fig:diag-U0-vsn-inf}
\end{figure*}

As it follows from numerical solutions of 
Eqs. (\ref{eqVeta})--(\ref{eqpeta}) (cf. also Fig. \ref{fig:THF_TKT_vsn}), for 
$J/t \gg 1$ ($t=0$, $0<n<2$) calculating the asymptotic expressions for $T_c$ one obtains:
\begin{equation}
k_{B} T_{c} = d J(n-1)\left[\ln \left( \frac{n}{2-n} \right)  \right]^{-1},  \nonumber  \label{Tct0} \\
\end{equation}
In this case, numerical solutions of Eq. (\ref{kBTc}) for the $T_{KT}$ 
(cf. Fig. \ref{fig:THF_TKT_vsn}) for $d=2$ is well approximated by $\rho _s(0)$, i.e., 
\begin{equation}
k_BT_{KT}\simeq Q \rho_{s}(0).  \label{C9}
\end{equation}
and using the expression for $\rho_s(0)= Jn(2-n)/2$ [obtained from Eqs. (\ref{gs}) and (\ref{eq:propTBpairs})] one obtains
\begin{equation}
k_{B} T_{KT} = Q J n (2-n) / 2.  \label{C10}
\end{equation}
As one sees, in the strong coupling limit, the ratio $T_{KT}/T_c$ substantially
depends on the band--filling. 
It is maximal for $n=1$ (cf. Fig. \ref{fig:TKTByTHF_vsn}): 
\begin{equation}
\frac{T_{KT}}{T_{c}}\left( n=1\right) = \frac Q {2d}  \label{C15}
\end{equation}
and decreases towards zero with the increase of $|n-1|$:
\begin{equation}
\frac{T_{KT}}{T_{c}}(n)  \simeq \frac{ Q}{4d} \frac{ 
n(2-n)}{(n-1)} \ln{ \left( \frac{n}{2-n} \right)}.  \label{C16}
\end{equation}

\section{Results for $d=\infty$  lattices}
\label{sec:appB}

For the case of $d=\infty$ lattice, we present only the most important results concerning the ground state 
and the critical temperature, which are concluded in the main text of the paper.
Let us stress again that for the considered model with intersite interactions only, the HFA becomes rigorous theory in the $d=\infty$ limit, and the presented here results are exact ones.

\begin{figure*}[t]
\centering
\includegraphics[width=\sizetwo]{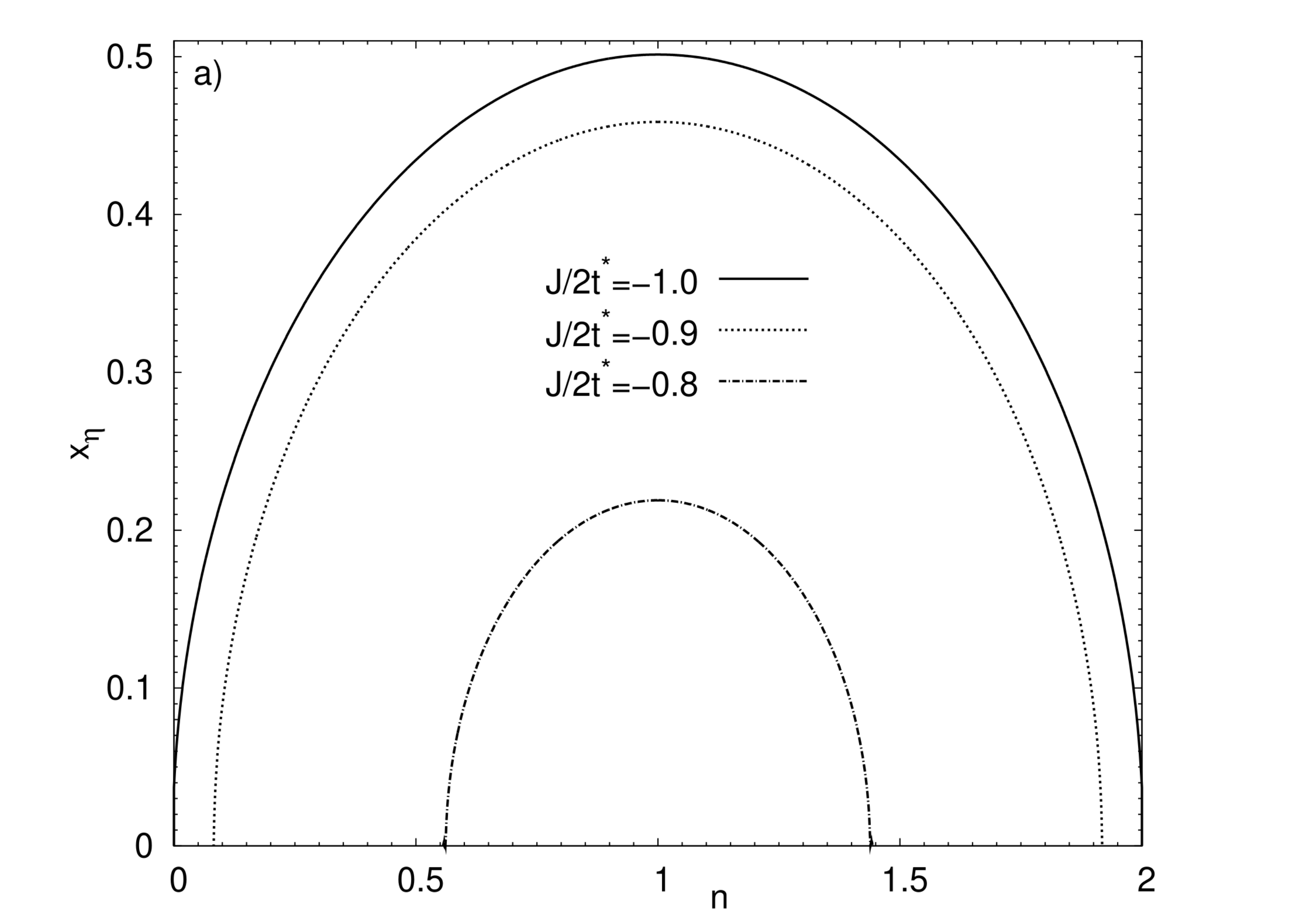}
\includegraphics[width=\sizetwo]{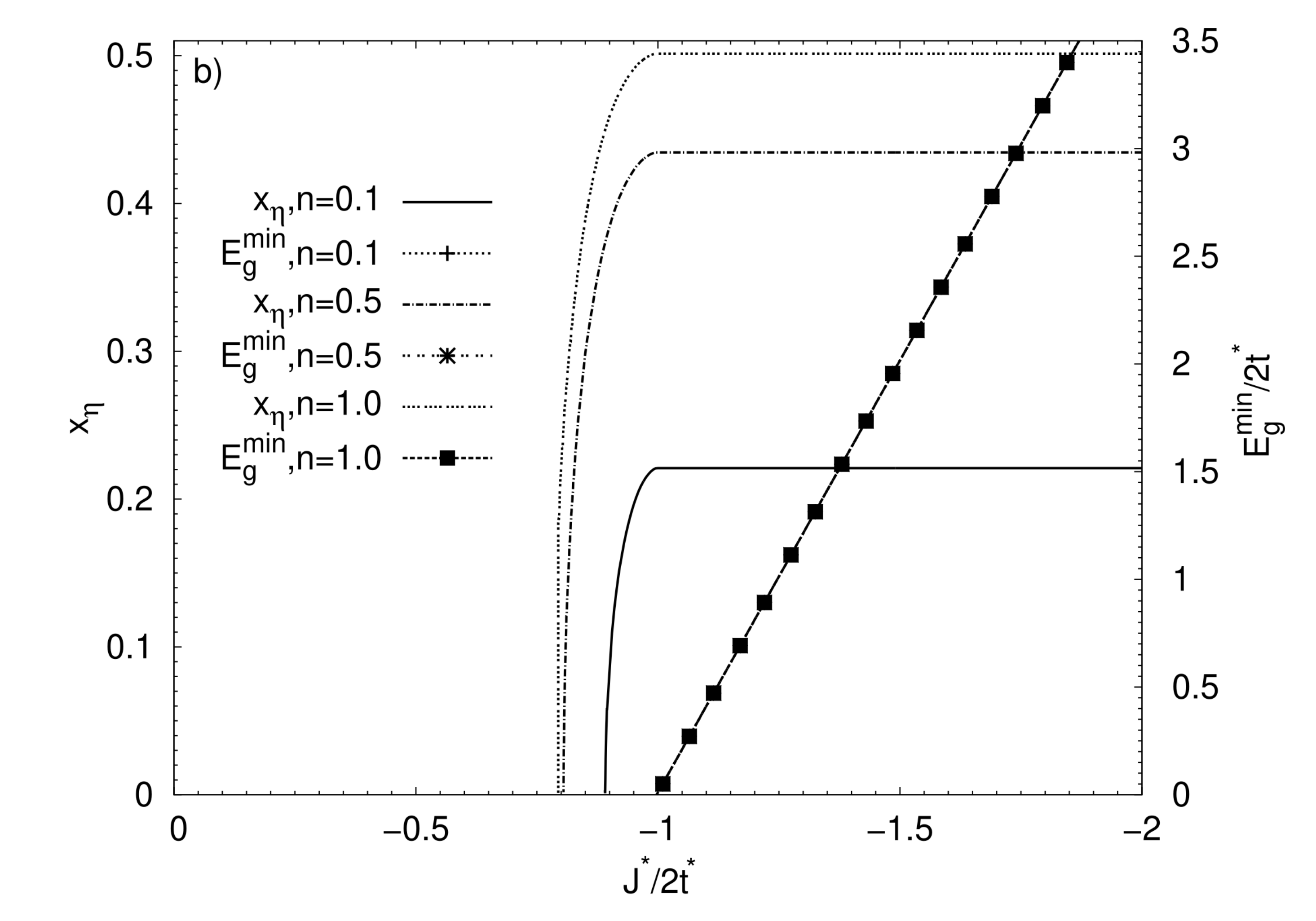}
\caption{
The ground state plots of the eta--pairing order parameter $x_{\eta}$:  
(a) as a function of concentration $n$ for several fixed values of $J/2t^{*}$: $J/2t^{*}=-0.8$, $J/2t^{*}=-0.9$, and $J/2t^{*}=-1$;
(b) as function of interaction $J$  for several fixed values of $n$: $n=1.0$, $n=0.5$, and $n=0.1$. 
The $E_g^{min}$ versus $J$ lines are denoted with black squares ($\blacksquare$).
They are the same for all $n$. 
The semi-elliptic density of states is used.
}\label{fig:x-Eg-vsn-inf}
\end{figure*}

\begin{figure*}[t]
\centering
\includegraphics[width=\sizetwo]{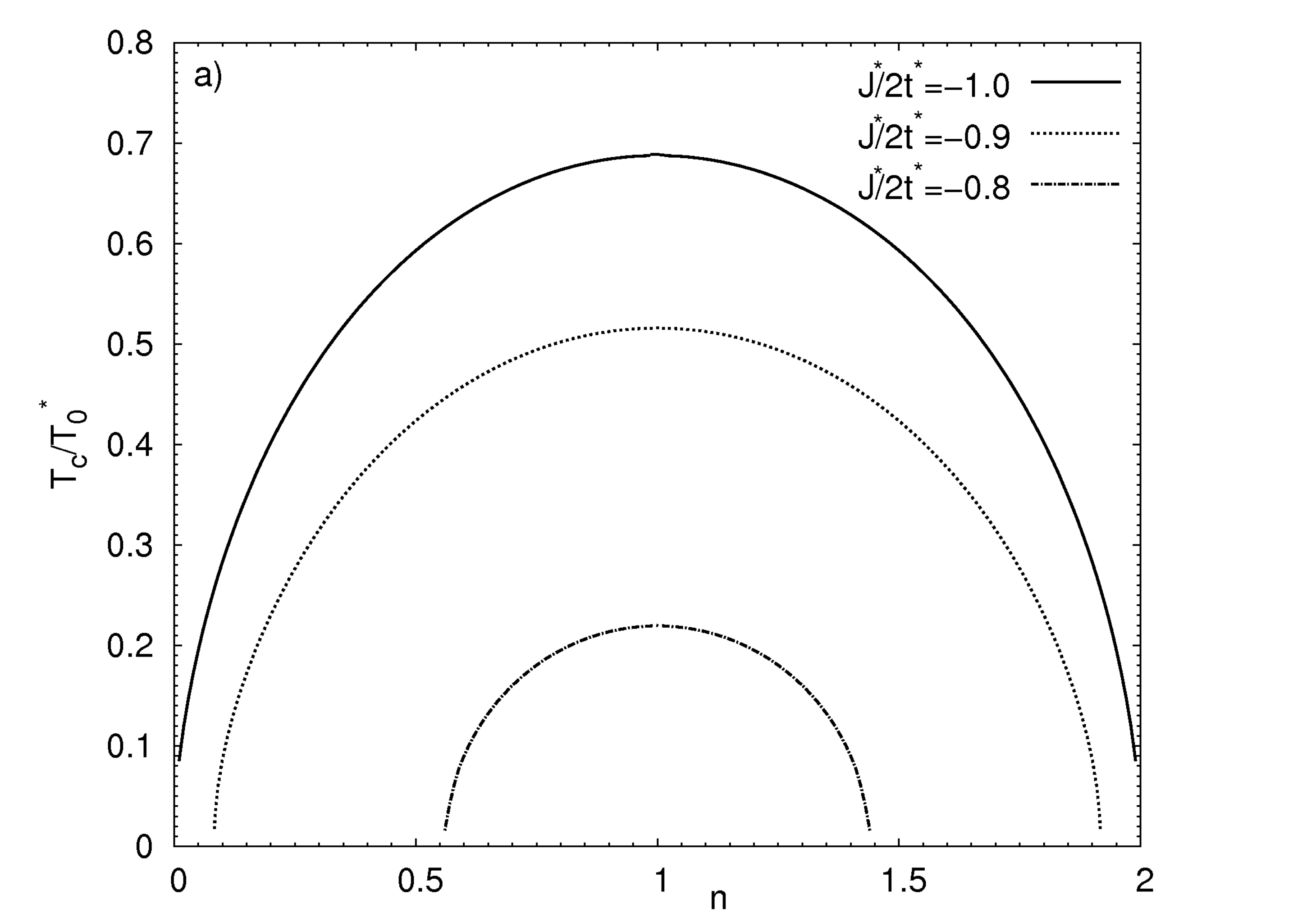}
\includegraphics[width=\sizetwo]{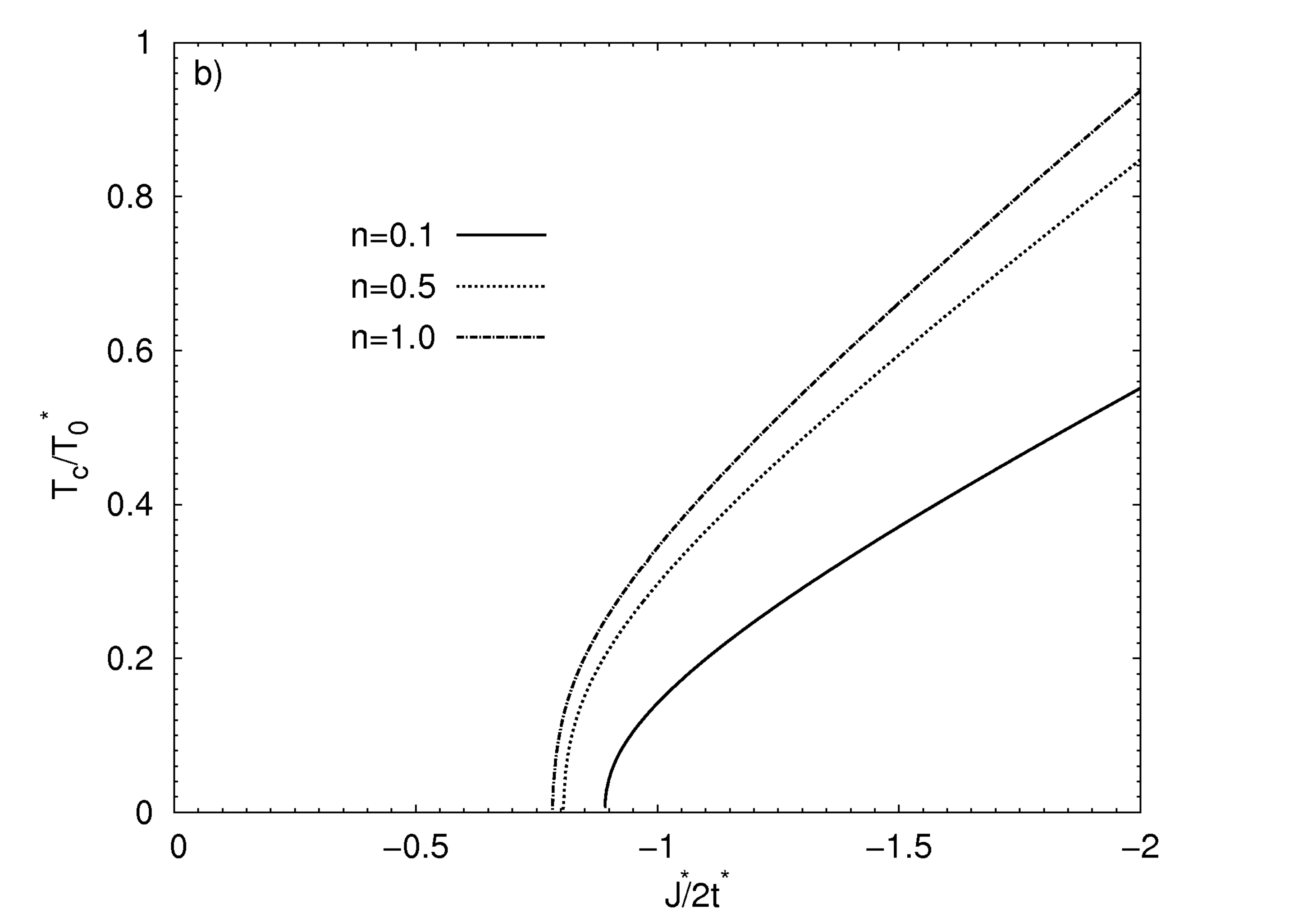}
\caption{
(a) Concentration $n$ and (b) interaction $J$ dependencies of the critical temperature $T_{c}/T_{0}$ ($T_0^{*}=2t^{*}/k_B$, the semi-elliptic density of states used). 
Curves on panel (a) are plotted for fixed values of $J^{*}/2t^{*}$: $J^{*}/2t^{*}=-0.8$, $J^{*}/2t^{*}=-0.9$, and $J^{*}/2t^{*}=-1.0$, while lines on panel (b) -- for fixed $n$: $n=0.1$, $n=0.5$, and $n=1$.
}\label{fig:THF-vsn-inf}
\end{figure*}

The plots of the ground state phase boundaries between the eta--phase and the N state obtained for semi-elliptic and Gaussian densities of states (DOS) are shown in Fig. \ref{fig:diag-U0-vsn-inf}, whereas, in Figs. \ref{fig:x-Eg-vsn-inf} and \ref{fig:THF-vsn-inf} we present, respectively, the eta--pairing order parameter $x_{\eta}$, energy gap $E_{g}^{min}$ and the critical temperature $T_{c}$ as a function of $n$ and $J^{*}$. The superconducting characteristics are plotted for semi-elliptic DOS.

As we find in Figs. \ref{fig:diag-U0-vsn} and \ref{fig:diag-U0-vsn-inf}, with increasing $|1-n|$,  the ground state phase boundaries between the eta--pairing and the N state are shifted towards higher values of $|J|$.
Thus, in a certain range of interaction parameter $J$ the transition from the superconducting eta--phase to the N state can be realized by changing the electron density.
In particular, the strongest $n$-dependence is observed in the case of SQ lattice, where at $n=1$ due to the van Hove singularity $J_c=0$ and  the eta--pairing phase can be stable for any $J<0$. 

In the following, we recapitulate the main features that differentiate the case of $d=\infty$ lattices from 
the other considered lattice structures.  
For $d=\infty$ lattice, except for the ranges close to the limits of empty or fully occupied band, 
$J_{c}$ weakly depends on $n$. 
This is in contrast to the case of SC structure, where is constant in analogous range of $n$ and to SQ lattice, where $J_{c}$ significantly depends on $n$ within the entire range of $n$.
In the system with NN hopping only, for the symmetric DOS structures considered in this work (except for the $d=\infty$ lattices), $\left| J_{c1}\right| =2t$ for any $n$.

In the weak eta--pairing phase, analogously as in cases of SQ and SC lattice, for $d=\infty$ lattice with semi-elliptic DOS for $\left| J_c\right| < \left| J\right| <\left| J_{c1}\right|$: $E_g^{min }<0$ and $x_{\eta} <x_{\eta} ^{max }$ (in the strong eta--phase $x_{\eta}=x_\eta ^{max }$). 
The weak eta--phase is stable only within a restricted range of concentration, and the range shrinks with decreasing $|J|$ and disappears for $J\rightarrow J_{c}$. 
On the other hand, in the case of  $d=\infty$ lattice with Gaussian DOS, with decreasing $\left| J\right|$, $x_{\eta}$ monotonically decreases and the range of $n$ occupied by eta--phase  shrinks [Fig. \ref{fig:diag-U0-vsn-inf}(b)] and  finally vanishes at $J_{c}$ (formally, in this case only the weak eta-phase occurs).

In Fig. \ref{fig:THF-vsn-inf} we plot the evolution of the $T_c$ versus $n$  and versus $J$ for a few fixed values of $J$ and $n$ (cf. Figs. \ref{fig:THF-vsn-sq} and \ref{fig:THF-vsn-sc} for the SQ and SC lattices, respectively).
In the weak eta-pairing phase, analogously to the $x_{\eta}$, the $T_{c}$ is restricted to a limited range of $n$ and the range gradually vanishes with $J\rightarrow J_{c}$. 
In the strong coupling limit, for the $d=\infty$, $T_{c} \sim J$ as for the other considered lattices.

\section{Densities of states used in the numerical calculations}
\label{sec:appC} 

In this work, the intersite hopping $t$ is restricted to nearest--neighbors. 
In such a case, density of states for the SQ lattice ($d=2$) is given by
\begin{equation}
D(\epsilon )_{2D}=\frac{1}{2t\pi ^2}K\left[ 1-\left( \frac{\epsilon}{4t} \right) ^2\right] ,
\end{equation}
if $|\epsilon /4t|<1$ and zero otherwise, where $K(x)$ denotes the complete elliptic integral of the first kind.
For $d=3$ (i.e., the SC lattice) an analytic approximation of $D(\epsilon )$ is used as calculated numerically in Ref.~\cite{Jelitto-69}.
For this lattice, also full numerical integration over the first Brillouin zone was performed. 

For the infinite dimensional lattices ($d=\infty $), we consider two different densities of states, namely:
(i) the semi-elliptic density of states in the form of
\begin{equation}\label{eq:dos.Bethe}
D(\epsilon )_{\infty D-B}=\frac{1}{2\pi (t^*)^2} \sqrt{4(t^*)^{2}-\epsilon^{2}} \quad \textrm{for} \quad |\epsilon|\leq 2t^*,\quad
\end{equation}
and  $D(\epsilon )_{\infty D-B}= 0$ for  $|\epsilon|>2t^*$, which is the DOS for the Bethe lattice (with $z\rightarrow \infty$), as well as 
(ii) the Gaussian density of states defined by
\begin{equation}
D(\epsilon )_{\infty D-G}=\frac{1}{t^{*}\sqrt{2\pi }}\exp \left[ -\frac{1}{2} \left( \frac{\epsilon}{t^{*}}\right) ^2\right],
\end{equation}
which is the DOS of the hypercubic lattice in the $d\rightarrow \infty$ limit. 
For $d=\infty $ the renormalized parameters are $t^{*}=t\sqrt{2d}$ and $J^{*}=Jd$ (and $z=2d$ for hipercubic lattices) \cite{Robaszkiewicz-99,czart-11}. 
In this limit the Fock term is irrelevant because the effective width of the band $B_{eff}=4 \widetilde{t} d=2t^{*} \sqrt{2d}+4J^{*}p/d$ and the second term vanishes for $d\rightarrow \infty$.
Note that Eq. (\ref{eq:dos.Bethe}) can be also considered as an approximation of the DOS for the SC lattice.

\subsection*{Declaration of competing interest}
The authors declare that they have no known competing financial interests or personal relationships that could have appeared to influence the work reported in this paper. The funders had no role in the design of the study; in the collection, analyses, or interpretation of data; in the writing of the manuscript, or in the decision to publish the results.

\subsection*{Acknowledgments}
We thank  Tomasz Kostyrko for very helpful discussions. 
K. J. K. acknowledges the support from the National Science Centre (NCN, Poland) under Grant SONATINA 1 no. UMO-2017/24/C/ST3/00276.
K. J. K. appreciates also  founding in the frame of a scholarship of the Minister of Science and Higher Education (Poland) for outstanding young scientists (2019 edition, no. 821/STYP/14/2019).

\subsection*{CRediT authorship contribution statement}
Wojciech R. Czart: Conceptualization, Methodology, Software, Formal analysis, Investigation, Resources, Data curation, Writing - original draft, Visualization. Konrad J. Kapcia: Validation, Formal analysis, Writing - original draft, Writing - review \& editing, Supervision, Project administration. Roman Micnas: Software, Formal analysis, Investigation, Resources, Data curation, Writing - original draft. Stanisław Robaszkiewicz: Conceptualization, Methodology, Writing - original draft, Supervision.

\bibliography{PHmodel-four}

\end{document}